\documentclass[prl,article,twocolumn]{revtex4-1}
\usepackage{amsmath,color,url,graphicx}
\usepackage{fontawesome5}
\usepackage[normalem]{ulem}
\usepackage[colorlinks=true,allcolors=blue]{hyperref}
\usepackage[compat=1.1.0]{tikz-feynhand}
\usepackage{verbatim}

\tikzset{graviton/.style={decorate, decoration={snake, amplitude=.4mm, segment length=1.5mm, pre length=.5mm, post length=.5mm}, double}}
\tikzset{inflaton/.style={thick, black, dashed}}
\tikzset{reheaton/.style={orange, dashed}}
\tikzset{fermion/.style={thick, black, postaction={decorate},decoration={markings,mark=at position 0.6 with {\arrow{stealth}}}}}

\newcommand{\Trh}{T_{\rm rh}}
\newcommand{\arh}{a_{\rm rh}}

\newcommand{\be}{\begin{equation}}
\newcommand{\ee}{\end{equation}}

\newcommand{\bea}{\begin{eqnarray}}
\newcommand{\eea}{\end{eqnarray}}
\newcommand{\nn}{\nonumber}

\begin{document}

\title{Irreducible Graviton Floor from Reheating}

\author{James M. Cline}
\email{jcline@physics.mcgill.ca}
\affiliation{McGill University Department of Physics \& Trottier Space Institute\\
3600 Rue University, Montr\'eal, QC, H3A 2T8, Canada}

\author{Yong Xu}
\email{yong.xu6@mcgill.ca}
\affiliation{McGill University Department of Physics \& Trottier Space Institute\\
3600 Rue University, Montr\'eal, QC, H3A 2T8, Canada}

\date{\today}
\begin{abstract}
Inflaton decay inevitably emits gravitons through bremsstrahlung during reheating.  
We show that the soft part of this emission amplitude, fixed by Weinberg's soft-graviton theorem, becomes an irreducible stochastic gravitational-wave (GW) background after accounting for cosmological evolution. 
The theorem fixes the infrared branch of the spectrum, $\Omega_{\rm GW}\propto f$, independently of the microscopic operator responsible for inflaton decay, while the normalization is controlled by the hard inflaton decay rate and by a phase-space factor. 
We carry this out for inflaton $n$-body decays, including the phase-space integrals,
finding that the maximum of the spectrum scales as $2/n$ relative to the $n=2$ case.
The signal can reach
$\Omega_{\rm GW}h^2\sim \mathcal O(10^{-17})$
at frequencies above the GHz scale. 
This predicts a stochastic graviton floor from perturbative reheating: a larger signal would require either other processes beyond perturbative bremsstrahlung or inflationary scenarios beyond conventional single-field slow roll.
\end{abstract}
\maketitle

{\bf Introduction.---}
Cosmic inflation gives a successful account of the large-scale homogeneity,
isotropy, and flatness of the Universe, and of the primordial perturbations
that seeded structure formation~\cite{Starobinsky:1980te,Guth:1980zm,
Linde:1981mu,Albrecht:1982wi}.  After inflation ends, however, the energy
stored in the inflaton condensate must be converted into ordinary
relativistic degrees of freedom.  This transition, known as reheating,
initiates the hot Big Bang and fixes the thermal history prior to Big Bang
nucleosynthesis~\cite{Allahverdi:2010xz,Amin:2014eta}.  Yet the microphysics
of reheating remains largely unconstrained, because the particles produced
during this epoch subsequently scatter, thermalize, and erase much of the
information about their origin.  Identifying observables that retain direct
memory of reheating therefore remains a central challenge in early-Universe
cosmology.

Gravitons are exceptional messengers of this epoch.  Once produced, they
propagate essentially freely and preserve information about their source.
During reheating, one source is unavoidable: graviton bremsstrahlung
accompanying inflaton decay.  Whenever the inflaton decays into lighter
particles, gravitational radiation will be emitted.  Previous studies have
mainly considered inflaton decays via nonderivative trilinear couplings
~\cite{Nakayama:2018ptw,Barman:2023ymn,Barman:2023rpg,Kanemura:2023pnv,
Bernal:2023wus,Tokareva:2023mrt,Xu:2024fjl,Bernal:2025lxp,Xu:2025wjq}.
A basic question is then the extent to which 
the resulting gravitational wave (GW) signal is determined by the universal coupling of gravity to energy-momentum.

We note that Weinberg's soft-graviton theorem~\cite{Weinberg:1965nx}
provides precisely such an operator-independent infrared pediction.  At the amplitude
level, the theorem states that soft graviton emission factorizes into the
hard decay process multiplied by a universal infrared pole. 
{For reheating through perturbative inflaton decay, we show that this factorized soft emission becomes, after phase-space integration and cosmological evolution, a perturbative baseline for the stochastic GW background. We refer to it as an irreducible graviton floor because it is fixed by the universal gravitational coupling once the hard decay rate is specified.}
The infrared branch is fixed by soft-graviton kinematics,
$\Omega_{\rm GW}\propto f$, independently of the microscopic decay operator, while the hard part of the spectrum depends only on the multiplicity of decay products.
In the following, we will show how the soft part of the spectrum transitions to the hard part, which quickly reaches a maximum and thereafter falls off, and how the
spectra scale inversely to the multiplicity of the inflaton decay products, as could be anticipated by consideration of  classical GW emission.

{\bf Setup and soft factorization.---}
We consider reheating in which the inflaton $\phi$
decays into light bosonic degrees of freedom $\varphi$, which may be
identified, for example, with the Higgs degree of freedom. A useful
set of benchmark interactions is
\begin{align}
\mathcal L_{\rm int}
=
-\frac{\mu}{2}\phi\varphi^2
-\frac{1}{2\Lambda}\phi(\partial_\mu\varphi)(\partial^\mu\varphi) 
-\sum_{n\geq 3}\frac{\lambda_n}{n!}\phi\varphi^n\,.
\label{eq:L_int}
\end{align}
The first two operators represent nonderivative and derivative
two-body decays, while the $\lambda_n$ operators provide models for decays into $n$ scalar final-state particles.  In
the following we denote the corresponding hard decay rate by
$\Gamma_{\phi\to n\varphi}$ and assume that one of the channels dominates
the reheating dynamics. 

Other interactions like $\phi^2\varphi^2$ can in principle
transfer inflaton energy through scatterings rather than decays.  This
channel is inefficient for converting the inflaton energy, since the
scattering rate redshifts rapidly and falls below the Hubble rate before
the inflaton energy is depleted, potentially leaving a remnant inflaton
population~\cite{Kofman:1994rk,Hooper:2018buz}.  This
is one motivation for assuming that reheating occurs through inflaton decays.

The trilinear coupling  $\mu\phi\varphi^2$ can
 trigger tachyonic production when $\phi$ oscillates through negative
values, but vacuum stability requires a self-interaction
$\lambda'\varphi^4$ with $\lambda'\geq \mu^2/(2m_\phi^2)$, which raises
the effective mass of $\varphi$ and suppresses the instability
~\cite{Dufaux:2006ee}.  Efficient tachyonic transfer is then restricted
to a narrow region near the stability bound, and is strongly suppressed
for sizable self-couplings $\lambda'$ ~\cite{Dufaux:2006ee,Fan:2021otj}.  

During reheating the inflaton condensate behaves effectively as pressureless matter, leading to matter-dominated behavior with Hubble expansion rate $H\simeq 2/(3t)$. 
Reheating ends when the decay rate becomes comparable, $H(T_{\rm rh})\simeq 2\Gamma_\phi/3$, giving the reheat temperature
\begin{equation}
T_{\rm rh}
\simeq
\sqrt{\frac{2}{\pi}}
\left(\frac{10}{g_*}\right)^{1/4}
\sqrt{M_P\Gamma_\phi}\,,
\label{eq:Trh}
\end{equation}
where $g_*$ is the number of relativistic degrees of freedom.

Gravitons are unavoidably produced from inflaton decay through bremsstrahlung. We now turn to the infrared structure of graviton emission. 
Consider a process with external momenta $\{p_i\}$  and amplitude $\mathcal M_n(p_i)$. 
The emission of an additional graviton with 4-momentum  $\omega^\mu=(E_\omega,\vec\omega)\ll \max\{p_i\}$ 
and polarization tensor $\epsilon_{\mu\nu}(\omega)$ satisfies Weinberg's soft-graviton theorem~\cite{Weinberg:1965nx},
\begin{equation}
\mathcal M_{n+1}(p_i;\omega)
\simeq
\kappa\,\epsilon_{\mu\nu}(\omega)
\sum_i \eta_i
\frac{p_i^\mu p_i^\nu}{p_i\cdot\omega}\,
\mathcal M_n(p_i),
\label{eq:soft_graviton_theorem}
\end{equation}
at leading order in the soft expansion (in powers of $\omega$). 
Here $\kappa=\sqrt{32\pi G}=2/M_P$ in our conventions, and $\eta_i=+1$ ($-1$) for outgoing (incoming) particles. 
Eq.~\eqref{eq:soft_graviton_theorem} follows from the gravitational Ward identity associated with diffeomorphism invariance and is independent of the detailed interaction responsible for the hard process.

Squaring Eq.~\eqref{eq:soft_graviton_theorem} and summing over physical transverse--traceless graviton polarizations gives
$|\mathcal M_{n+1}|^2
\simeq
M_P^{-2}
S(\omega;\{p_i\})
|\mathcal M_n|^2$,
where $S(\omega;\{p_i\})$ is a soft kinematic factor with the characteristic infrared behavior $S\propto 1/E_\omega^2$.
Applying this result to inflaton decay, the amplitude for
$\phi\to \varphi_1\cdots\varphi_n+h(\omega)$ has the infrared form
\begin{equation}
|\mathcal M_{\phi\to n\varphi+h}|^2
=
|\mathcal M_{\phi\to n\varphi}|^2
\frac{1}{M_P^2}
\frac{m_\phi^2}{2E_\omega^2}\,
\mathcal{F}_{\rm soft}^{(n)} + \mathcal{O}(E_\omega^{-1})\, ,
\label{eq:universal_soft_decay}
\end{equation}
where $\mathcal{F}_{\rm soft}^{(n)}$  is a dimensionless function of relative angles between the decay products (with $\mathcal{F}_{\rm soft}^{(2)}=1$ for the case $n=2$), which determines the normalization of the observable soft GW spectrum. 

We compute these ingredients explicitly for the benchmark operators in
Eq.~\eqref{eq:L_int}.  Different operators lead to distinct hard amplitudes
and graviton-emission diagrammatics.  For example, the derivative
interaction contains a nonvanishing contact graviton vertex,
$\phi h_{\mu\nu}(\partial^\mu\varphi)(\partial^\nu\varphi)$, whereas the
nonderivative interaction gives only a trace contact term,
$h\phi\varphi^2$, which vanishes for an on-shell transverse--traceless
graviton.  Higher-multiplicity decays introduce additional external-leg
emission channels and a nontrivial phase-space average over the hard final
state.  These differences modify the residue of the $1/E_\omega^2$ pole and the hard decay
normalization, but not the universal infrared scaling. 

Eq.~\eqref{eq:universal_soft_decay} is the reheating-specific form of
Eq.~\eqref{eq:soft_graviton_theorem}.  It separates the hard squared
amplitude from the finite soft residue, making explicit the ingredients
that enter the phase-space average and the Boltzmann collision term.  This
organization is what allows the amplitude-level soft factorization to be
mapped onto the irreducible GW floor derived below.

{\bf From amplitudes to gravitational waves.---}
We next derive the observable GW spectrum arising from the decay amplitudes. 
The relevant quantity is the graviton phase-space distribution $f_h(t,p_h)$, whose evolution obeys
\begin{equation}
\frac{\partial f_h}{\partial t}
-
H p_h \frac{\partial f_h}{\partial p_h}
=
\mathcal C\, ,
\label{eq:Boltzmann}
\end{equation}
where $p_h\equiv |{\bf p}_h|$ is the physical graviton momentum and $\mathcal C$ is the collision term. 
For a generic process producing one graviton with four-momentum $\omega^\mu$ and energy $E_\omega$,
\bea
\mathcal C(p_h)
&=&
\frac{{\cal S}}{2E_\omega}
\int d\Pi_{\rm in}\,d\Pi_{\rm out}\,
f_{\rm in}|\mathcal M|^2
(2\pi)^4\nn\\
&\,& \qquad \times\
\delta^{(4)}(P_{\rm in}-P_{\rm out}-\omega),
\label{eq:collision_general}
\eea
where $d\Pi_i=d^3{\bf p}_i/[(2\pi)^3 2E_i]$ is the Lorentz-invariant phase-space measure. 
Here $d\Pi_{\rm in}$ and $d\Pi_{\rm out}$ denote the products of these measures over all non-graviton initial and final particles, $P_{\rm in}$ and $P_{\rm out}$ are their total four-momenta, $f_{\rm in}$ is the product of the initial-state distribution functions, and ${\cal S}$ is the symmetry factor for identical particles in the final state.

\begin{figure}[!t]
\def\sepf{0.6}
\centering
\includegraphics[scale=\sepf]{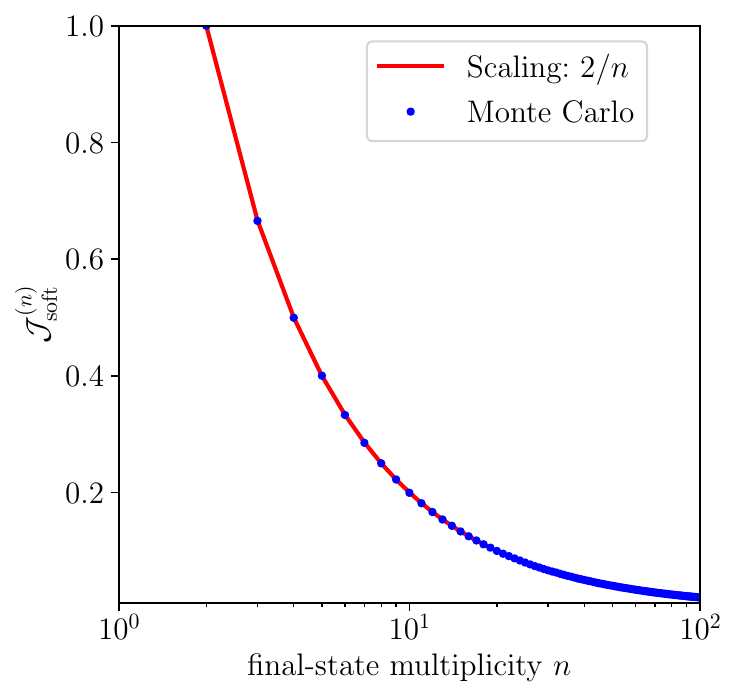}
\caption{Multiplicity dependence of  $\mathcal J_{\rm soft}^{(n)}$.
The blue points show the phase-space average of the polarization-summed
Weinberg soft factor, normalized to the two-body result, for massless
$n$-body final states.  Each point is obtained with $3\times10^5$
Monte Carlo events.  The agreement with $2/n$ shows that increasing
final-state multiplicity suppresses the transverse-traceless anisotropy
of the hard final-state stress tensor, so that the two-body channel gives
the maximal perturbative soft-graviton source.
}
\label{fig:Jn}
\end{figure}

Combining Eqs.~\eqref{eq:universal_soft_decay} and
\eqref{eq:collision_general}, the soft graviton source from the decay
$\phi\to n\varphi+h$ takes the form
\begin{equation}
\mathcal C^{(n)}_{\rm soft}
\simeq
\frac{n_\phi\Gamma_{\phi\to n\varphi}}{M_P^2 m_\phi}
\frac{2}{x^3}\,\mathcal J_{\rm soft}^{(n)},
\qquad
x\equiv \frac{2E_\omega}{m_\phi}\, .
\label{eq:C_soft_universal}
\end{equation}
{Eq.~\eqref{eq:C_soft_universal} is the  manifestation of soft graviton theorem at level of collision-term.} Here $n_\phi$ is the inflaton number density,
$\Gamma_{\phi\to n\varphi}$ is the corresponding decay rate with no gravitons, and
$\mathcal J_{\rm soft}^{(n)}$ is the phase-space average of $\mathcal F_{\rm soft}^{(n)}$ in
Eq.~\eqref{eq:universal_soft_decay}. The factor $x^{-3}$ comes from
the leading infrared behavior, $|\mathcal M|^2\propto E_\omega^{-2}$, together
with the graviton phase-space factor $1/(2E_\omega)$. All dependence on
the final-state multiplicity is therefore contained in  $\mathcal J_{\rm soft}^{(n)}$.
For two-body decays the recoil kinematics is trivial,
giving $\mathcal{J}_{\rm soft}^{(2)}=1$. For higher multiplicities, we numerically determine the phase space average to be consistent with 
\begin{equation}\label{eq:Jsoft_n}
\mathcal J_{\rm soft}^{(n)}=\frac{2}{n}\,.
\end{equation}
as depicted by the red solid line in Fig.~\ref{fig:Jn}. 
The technical construction of this phase-space average, together with its
classical quadrupole interpretation (Fig.~\ref{fig:quadrupole_intuition}), is given in
Sec.~\ref{sec:TT_ quadrupole}. 

The physical interpretation of Eq.~\eqref{eq:Jsoft_n} is transparent: as $n$ increases, the hard
energy--momentum is shared among more final-state particles and becomes more
isotropically distributed. Since gravitons couple to the transverse-traceless
part of the stress tensor, this isotropization suppresses the soft-graviton
normalization while leaving the infrared scaling unchanged.

The next task is to solve the Boltzmann equation for the present-day graviton phase-space distribution. For this we introduce the comoving graviton momentum $\tilde p_h\equiv a p_h$. 
The Boltzmann equation becomes $d f_h(\tilde p_h)/dt=\mathcal C(\tilde p_h)$, with solution
\begin{equation}
f_h(a)
=
\int_{a_I}^{a}
\frac{\mathcal C(\tilde p_h)}
{a' H(a')}
\,da' ,
\label{eq:fh_solution}
\end{equation}
where $a_I$ is the scale factor at the onset of reheating. 
Thereafter, the inflaton condensate behaves as pressureless matter, so that $H(a)\simeq H_I(a/a_I)^{-3/2}$ and $n_\phi(a)\simeq 3M_P^2H^2(a)/m_\phi$. 
Using Eq.~\eqref{eq:C_soft_universal} in Eq.~\eqref{eq:fh_solution}, and evaluating at the end of reheating, $a=a_{\rm rh}$, gives
\begin{equation}
f^{(n)}_{h,\text{soft}}(a_{\rm rh})
\simeq
\frac{8\,\Gamma_\phi^2}{3m_\phi^2}
\left(\frac{m_\phi}{2p_h}\right)^3
\mathcal J_{\rm soft}^{(n)}
\left[1+\mathcal{O}\left(\frac{2p_h}{m_\phi}\right)\right] .
\label{eq:fh_soft_universal}
\end{equation}
At this point, the inflaton abundance has been exponentially depleted and graviton production becomes negligible. 
The distribution therefore freezes in and subsequently evolves only through cosmological redshift. 

The GW spectrum at present follows from
$\Omega_{\rm GW}(f)=\rho_c^{-1}d\rho_{\rm GW}/d\ln p_h
=16\pi^2 f^4 f_h(a_0,2\pi f)/\rho_c$,
where $p_h(a_0)=2\pi f$, $a_0$ is the scale factor today, and
$\rho_c\simeq 1.05\times10^{-5}h^2\,{\rm GeV/cm^3}$ is the critical density~\cite{ParticleDataGroup:2024cfk}. 
The graviton energy density $\rho_{\rm GW}$ includes both polarizations. 
Using Eq.~\eqref{eq:fh_soft_universal}, and taking the redshift from the end of reheating to today into account, the soft branch of the spectrum is
\begin{align}
\Omega^{\rm soft}_{\rm GW}h^2
\simeq
4.1\cdot 10^{-18}\,
\mathcal J_{\rm soft}^{(n)}
\left(\frac{\Gamma_\phi}{2.5\cdot 10^8\,{\rm GeV}}\right)^{1/2}\nonumber \\
\left(\frac{m_\phi}{10^{13}\,{\rm GeV}}\right)
\left(\frac{f}{10^{10}\,{\rm Hz}}\right),
\label{eq:Omega_soft_Gamma}
\end{align}
valid below the redshifted kinematic endpoint (derived below). 
Equivalently, using Eq.~\eqref{eq:Trh},
\bea
\Omega^{\rm soft}_{\rm GW}h^2
&\simeq&
3.8\cdot 10^{-18}\,
\mathcal J_{\rm soft}^{(n)}
\left(\frac{T_{\rm rh}}{10^{13}\,{\rm GeV}}\right)
\nn\\
&\times&
\left(\frac{m_\phi}{10^{13}\,{\rm GeV}}\right)\left(\frac{f}{10^{10}\,{\rm Hz}}\right).
\label{eq:Omega_soft_Trh}
\eea
Eqs.~\eqref{eq:fh_soft_universal}--\eqref{eq:Omega_soft_Trh}
are valid in the infrared regime, below the cutoff determined below.  They
promote Weinberg's soft-graviton theorem to a cosmological prediction:
the theorem fixes the low-frequency amplitude-level source, while the
Boltzmann evolution through reheating and subsequent redshift turn this
microscopic source into the present-day GW spectrum.

The microscopic physics enters only through the decay rate, equivalently $T_{\rm rh}$, and the phase-space average $\mathcal J_{\rm soft}^{(n)}$ of the differential decay rate residue. 
Once these quantities are known, the low-frequency shape and amplitude are fixed. Eq.~\eqref{eq:Omega_soft_Trh} also applies to fermion and vector final states; see Refs.~\cite{Barman:2023ymn,Xu:2025wjq}.

{\bf Graviton floor.---}
The low-energy spectrum terminates at frequencies above the redshifted inflaton mass.
At production the graviton energy is bounded by $E_\omega\leq m_\phi/2$, giving the present-day  endpoint
\begin{align}
f_{\rm th}
&\simeq
\frac{m_\phi}{4\pi}
\frac{a_{\rm rh}}{a_0}
=
\frac{m_\phi}{4\pi}
\left(\frac{T_0}{T_{\rm rh}}\right)
\left[\frac{g_{*s}(T_0)}{g_{*s}(T_{\rm rh})}\right]^{1/3}
\nonumber\\
&\simeq
9\times10^9\,{\rm Hz}
\left(\frac{m_\phi}{10^{13}\,{\rm GeV}}\right)
\left(\frac{10^{13}\,{\rm GeV}}{T_{\rm rh}}\right)\,,
\label{eq:fth_model_independent}
\end{align}
where $a_0$ denotes the scale factor at present, and $g_{*s}$ denotes the effective number of entropy degrees of freedom. The spectrum turns over before this point; 
in the exact calculation we find  it peaks at frequency $f_{\rm peak}\sim \mathcal O(0.1)\, f_{\rm th}$, giving the maximum amplituden
\begin{equation}\label{eq:floor}
\Omega_{\rm GW}^{\rm floor}h^2
\simeq
\mathcal O(10^{-19})
\mathcal J_{\rm soft}^{(n)}
\left(\frac{m_\phi}{10^{13}\,{\rm GeV}}\right)^2 .
\end{equation}
Thus the magnitude of the perturbative reheating floor is controlled by $m_\phi$, while the peak frequency also depends upon the reheating temperature.

\begin{figure}[!t]
\def\sepf{0.6}
\centering
\includegraphics[scale=\sepf]{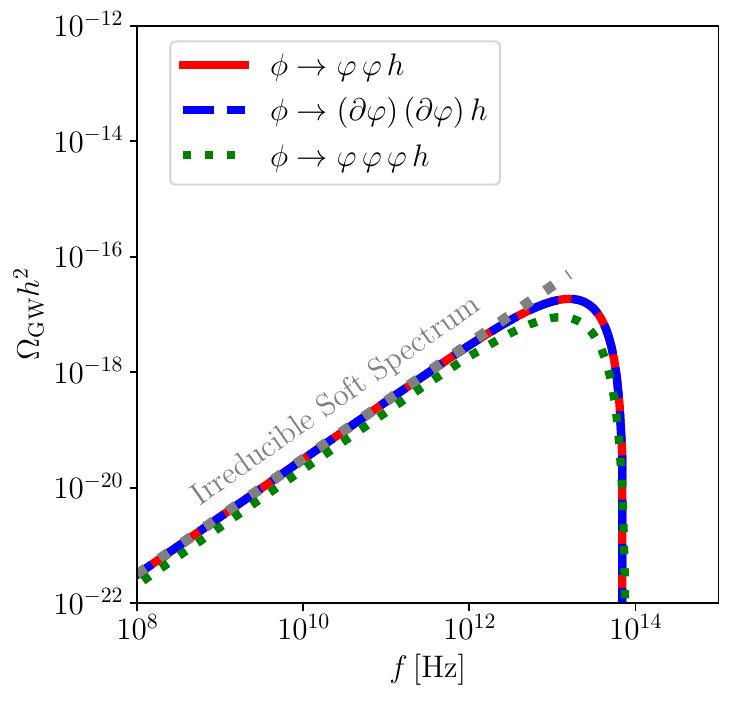}
\caption{Comparison of the GW spectra from the three benchmark reheating channels:
$\phi \to \varphi\varphi h$ through the nonderivative coupling (red solid),
$\phi \to (\partial\varphi)(\partial\varphi)h$ through the derivative coupling
(blue dashed), and $\phi \to \varphi\varphi\varphi h$ (green dotted), for
$m_\phi = 8.3\times 10^{13}~\mathrm{GeV}$ and $\Trh = 10^{10}~\mathrm{GeV}$.
The gray dotted line shows the infrared approximation in Eq.~\eqref{eq:Omega_soft_Trh}.}
\label{fig:GW_2body_3_body}
\end{figure}

Constructing the spectrum near its maximum requires the dependence on the emitted graviton energy. 
For this reason, one must compute the graviton-emission amplitudes
beyond the soft approximation in Eq.~\eqref{eq:soft_graviton_theorem}.
The relevant calculations are presented in Sec.~\ref{sec:2body} for
derivative and nonderivative two-body decays, and in
Sec.~\ref{sec:3body} for the three-body channel, including the
phase-space integral and the resulting GW spectrum beyond the soft limit. For two-body decays it can be done analytically, since specifying the graviton momentum fixes the recoiling two-particle kinematics. For
higher-multiplicity decays, the recoil system retains internal phase space, so the collision term becomes a multidimensional phase-space integral. We obtained the spectrum by first evaluating the collision term numerically  and then evolving 
the Boltzmann equation through reheating until the present.

Figure~\ref{fig:GW_2body_3_body} shows selected spectra beyond the soft
limit.  Once normalized to the same hard decay rate, the derivative and
nonderivative two-body spectra coincide over the full kinematic range.
The reason is that the graviton probes the total energy--momentum flow in
the decay.

After all 
diagrams are included, both operators describe the same
two-body energy--momentum flow, differing only by the hard decay strength;
this is shown in Sec.~\ref{sec:2body}.  The three-body spectrum
is obtained from the phase-space evaluation and is suppressed
relative to the two-body case, consistently with
$\mathcal J_{\rm soft}^{(3)}=2/3$.

We can further constrain the graviton floor amplitude
(\ref{eq:floor}) using generic expectations from  single-field inflation.  Namely, the inflaton mass should not exceed the Hubble rate during slow roll, $m_\phi\lesssim \sqrt{3}H_{\text{inf}}$, giving
\begin{align}\label{eq:floor_bound}
  \Omega^{\rm floor}_{\rm GW} h^2
  \lesssim
  \mathcal{O}(10^{-17})
  \left( \frac{\mathcal{P}}{2.1 \times 10^{-9}}\right)
  \left( \frac{r}{0.01} \right),
\end{align}
where $H_{\text{inf}}$ is the inflationary Hubble scale, $\mathcal P\simeq2.1\times10^{-9}$ \cite{Planck:2018vyg} denotes the scalar power spectrum amplitude, and $r$ is the tensor-to-scalar ratio, which is bounded by $r\leq 0.035$ from BICEP/Keck 2018 \cite{BICEP:2021xfz}. 

\begin{figure}[!t]
\def\sepf{0.6}
\centering
\includegraphics[scale=\sepf]{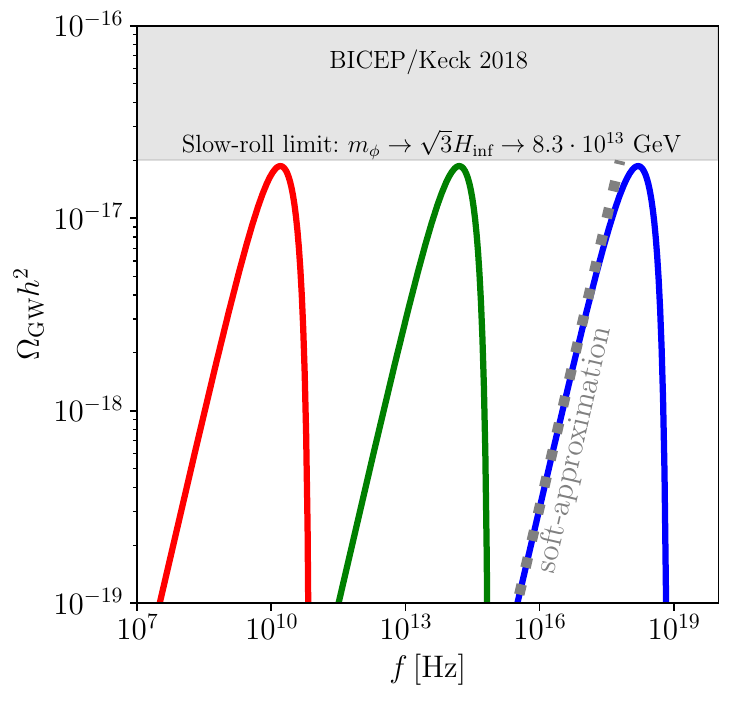}
\includegraphics[scale=\sepf]{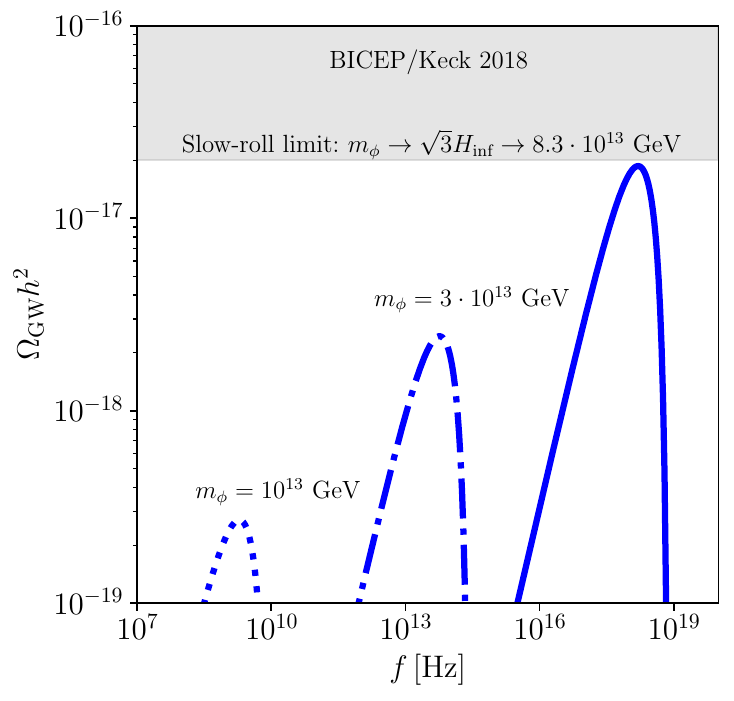}
\caption{Graviton floor from bremsstrahlung during reheating. The amplitude is controlled
primarily by $m_\phi$, while the peak frequency scales as $m_\phi/\Trh$. The gray dotted
line in the upper panel shows the infrared approximation, Eq.~\eqref{eq:Omega_soft_Trh},
valid for $f\ll f_{\rm th}$. The gray region is ruled out by limits on the tensor-to-scalar ratio $r$ \cite{BICEP:2021xfz}.}.
\label{fig:GW_floor}
\end{figure}

The criterion $m_\phi\lesssim\sqrt{3}H_{\text{inf}}$ is valid in
typical single-field inflation models, that obey the slow-roll condition $\eta \equiv M_P^2 V''(\phi_*)/V(\phi_*)\ll1$, where $\phi_*$ denotes the field value at horizon crossing. 
It implies $m_{\rm eff}^2\equiv |V''(\phi_*)|=3|\eta_*|H_{\text{inf}}^2<3H_{\text{inf}}^2$. 
It applies for popular models such as Starobinsky inflation~\cite{Starobinsky:1980te}, attractor inflation~\cite{Kallosh:2013hoa}, and polynomial inflation~\cite{Hodges:1989dw,Drees:2022aea}.
However it can be evaded in models where the inflation mass changes significantly between the slow-roll phase and reheating, in particular two-field models like hybrid inflation \cite{Linde:1993cn}.

Fig.~\ref{fig:GW_floor} shows the graviton floor for representative values of 
$\Trh$ and $m_\phi$. In both panels the colors distinguish the reheating
temperatures: red, green, and blue correspond to $\Trh=10^{13},\,10^9,\,10^5\,{\rm GeV}$, respectively. In the upper panel, the inflaton mass is fixed at the slow-roll-motivated value
$m_\phi\simeq8.3\times10^{13}\,{\rm GeV}$. In the lower panel, the line
style denotes the inflaton mass: dotted, dash-dotted, and solid correspond to $m_\phi=10^{13},\,3\times10^{13},\,8.3\times10^{13}\,{\rm GeV}$,
respectively. Lower $\Trh$ shifts the peak to higher frequency because the gravitons are redshifted less after production, while increasing $m_\phi$ mainly raises the amplitude.

The maximal signal of the perturbative graviton floor lies
below both proposed high-frequency GW sensitivities
\cite{Aggarwal:2025noe} and the integrated Big Bang nucleosynthesis bound
$\Omega_{\rm GW}h^2\lesssim 10^{-6}$
\cite{Maggiore:1999vm}, inferred from Planck 2018 constraints on dark
radiation~\cite{Planck:2018vyg}.  The significance of this result is
therefore not that the signal is close to detection, but that it gives a
rigorous ceiling for graviton bremsstrahlung from perturbative
reheating.  Within this regime, ordinary perturbative inflaton decay
cannot naturally generate a much larger high-frequency GW background.  A
detectable signal from the same epoch may point to physics
beyond the perturbative reheating baseline, such as nonperturbative
dynamics or inflationary/reheating scales outside the conventional
single-field slow-roll expectation.

Other GW sources can also arise during reheating, but they are physically
distinct from the perturbative bremsstrahlung floor derived here.  The
inflationary GW background from modes reentering the horizon during reheating is
negligible at the high frequencies considered in this work; 
see, e.g.,
Ref.~\cite{Xu:2025wjq}. Inflaton annihilations can also source gravitons; however,
the corresponding spectra are more strongly suppressed, scaling as
$1/M_P^4$~\cite{Ema:2020ggo}.  
Thermal GWs from the Standard Model plasma can
peak at frequencies of order $\mathcal O(100)\,{\rm GHz}$
\cite{Ghiglieri:2015nfa,Ghiglieri:2020mhm,Ringwald:2020ist,Bernal:2024jim,Xu:2024cey}.
Their amplitude is controlled by the plasma temperature and is strongly
suppressed for low reheating temperatures; away from the thermal peak,
the spectrum also decreases rapidly
\cite{Bernal:2024jim,Xu:2024cey}.  These backgrounds should therefore be regarded as separate components of the high-frequency GW budget. Our result does not attempt to exclude them; instead it isolates the contribution that is unavoidable assuming reheating proceeds through perturbative inflaton decay.

{\bf Conclusion.---}We have exploited Weinberg's soft-graviton theorem to derive a cosmological
prediction from reheating. The theorem fixes the amplitude-level infrared
source associated with perturbative decays. After matching this source to the Boltzmann collision term and evolving it through the reheating epoch, the universal soft structure becomes an irreducible stochastic gravitational wave floor. Beyond the infrared regime, the spectrum is obtained from the full emission amplitudes, which connect smoothly onto the soft branch.

Our prediction is largely insensitive to the microphysical details of the inflaton decays.
Different two-body decay operators do not
give different signals for equal hard decay rates.
For example, derivative interactions lead to the same prediction as nonderivative contact interactions.
  Higher-multiplicity
decays with $n$ final state particles are only mildly suppressed (by a factor of $2/n$) relative to the 
two-body ($n=2$) decay channel, which gives the maximum
signal.

For inflaton masses motivated by conventional single-field slow roll, the maximum amplitude is at most $\Omega_{\rm GW}h^2\sim 10^{-17}$ at GHz and higher
frequencies, below current and proposed sensitivities.  
Hence any future observation exceeding this floor, assuming technological advances that might give access to such high frequencies and small amplitudes \footnote{We recall Einstein's belief that gravitational waves would never be observable at all}, would be evidence for physical processes distinct from 
perturbative reheating after inflation. 

\textit{Acknowledgments.---}
This work was supported by the Natural Sciences and Engineering Research Council of Canada (NSERC).

\appendix

\setcounter{section}{0}
\setcounter{subsection}{0}
\setcounter{equation}{0}
\setcounter{figure}{0}
\setcounter{table}{0}

\renewcommand{\thesection}{S\arabic{section}}
\renewcommand{\thesubsection}{S\arabic{section}.\arabic{subsection}}
\renewcommand{\theequation}{S\arabic{section}.\arabic{equation}}
\renewcommand{\thefigure}{S\arabic{figure}}
\renewcommand{\thetable}{S\arabic{table}}

\newcommand{\suppsection}[1]{%
  \refstepcounter{section}%
  \setcounter{subsection}{0}%
  \setcounter{equation}{0}%
  \par\bigskip
  \begin{center}
  \textbf{\thesection.\ #1}
  \end{center}
  \medskip
}

\newcommand{\suppsubsection}[1]{%
  \refstepcounter{subsection}%
  \par\medskip
  \begin{center}
  \textbf{\thesubsection.\ #1}
  \end{center}
  \medskip
}

\begin{center}
{\bf Supplemental Material}\\
{\bf Irreducible Graviton Floor from Reheating}
\end{center}
In this Supplemental Material we provide the technical details supporting
the Letter.  Section~\ref{sec:Feynman_Rules} summarizes the Feynman rules.
Sections~\ref{sec:2body} and \ref{sec:3body} give the full tree-level
amplitudes for derivative and nonderivative two-body decays and for the
three-body channel, including the corresponding collision terms with multidimensional phase-space integral, and the
gravitational-wave spectra beyond the soft limit.  Section~\ref{sec:TT_ quadrupole}
presents the technical  setup as well as Monte Carlo  algorithm to compute  $\mathcal J_{\rm soft}^{(n)}$ for inflaton $n$-body decays, together
with a classical quadrupole interpretation.  The numerical codes used
for the phase-space integrations and spectrum calculations are publicly
available on GitHub~\href{https://github.com/yongxuDM/Soft-Graviton}{\faGithub}.

\suppsection{Feynman Rules}\label{sec:Feynman_Rules}
Here, we summarize the Feynman rules relevant for the computation of the matrix elements. From Eq.~\eqref{eq:L_int}, the momentum-space vertices are found to be
\begin{align}
V_{\varphi\varphi h}^{\mu\nu}(p,q)
&=
\frac{i}{M_P}\Big[
p^\mu q^\nu+p^\nu q^\mu-\eta^{\mu\nu}\big(p\cdot q-m_\varphi^2\big)
\Big],\label{eq:Feynman_Rules_1}
\\
V_{\phi\phi h}^{\mu\nu}(p,q)
&=
\frac{i}{M_P}\Big[
p^\mu q^\nu+p^\nu q^\mu-\eta^{\mu\nu}\big(p\cdot q-m_\phi^2\big)
\Big],\label{eq:Feynman_Rules_2}
\\
V_{\phi\varphi\varphi}(l,p,q)
&=
\frac{i}{\Lambda}\,(p\cdot q),\label{eq:Feynman_Rules_3}
\\
V_{\phi\varphi\varphi h}^{\mu\nu}(l,p,q,k)
&=
\frac{i}{\Lambda M_P}
\left[
\eta^{\mu\nu}(p\cdot q)-\big(p^\mu q^\nu+p^\nu q^\mu\big)
\right],\label{eq:Feynman_Rules_4}
\\
V_{\phi\varphi\varphi}^{\rm (ND)}(l,p,q)
&=
-i\mu,\label{eq:Feynman_Rules_5}
\\
V_{\phi\varphi\varphi \varphi }(l,p,q,r)
&=
-i\lambda.\label{eq:Feynman_Rules_6}
\end{align}
For vertices involving the graviton, all signs and prefactors have been carefully checked against the existing literature, e.g.~Ref.~\cite{Choi:1994ax}.

\suppsection{Graviton Bremsstrahlung from Inflaton $2$-body Decay: Derivative, Nonderivative Coupling and Complete Spectrum}\label{sec:2body}

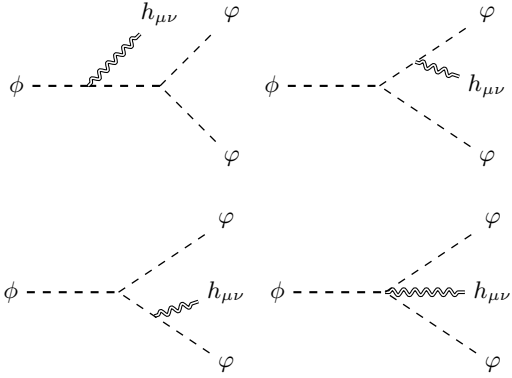
\begin{figure}[!ht]
\begin{tikzpicture}[scale=0.95]
    \begin{feynhand}
        \vertex (i) at (-1.5, 0) {$\phi$};
        \vertex (L) at (-0.5, 0);
        \vertex (R) at (0.5, 0);
        \vertex (3) at (1.5, 1) {$\varphi$};
        \vertex (4) at (1.5, -1) {$\varphi$};
        \vertex (h) at (0.5, 1) {$h_{\mu\nu}$};
        \vertex () at (0, -1.5) {};

        \propag [inflaton] (i) to (L);
        \propag [inflaton] (L) to (R);
        \propag [graviton] (L) to (h);
        \propag [scalar] (R) to (3);
        \propag [scalar] (4) to (R);
    \end{feynhand}
\end{tikzpicture}
\begin{tikzpicture}[scale=0.95]
    \begin{feynhand}
        \vertex (i) at (-1.5, 0) {$\phi$};
        \vertex (M) at (0, 0);
        \vertex (3) at (1.5, 1) {$\varphi$};
        \vertex (4) at (1.5, -1) {$\varphi$};
        \vertex (h) at (1.5, 0) {$h_{\mu\nu}$};
        \vertex () at (0, -1.5) {};

        \propag [inflaton] (i) to (M);
        \propag [graviton] (0.5,0.333) to (h);
        \propag [scalar] (M) to (3);
        \propag [scalar] (4) to (M);
    \end{feynhand}
\end{tikzpicture}
\begin{tikzpicture}[scale=0.95]
    \begin{feynhand}
        \vertex (i) at (-1.5, 0) {$\phi$};
        \vertex (M) at (0, 0);
        \vertex (3) at (1.5, 1) {$\varphi$};
        \vertex (4) at (1.5, -1) {$\varphi$};
        \vertex (h) at (1.5, 0) {$h_{\mu\nu}$};
        \vertex () at (0, -1.5) {};
        \propag [inflaton] (i) to (M);
        \propag [graviton] (0.5,-0.333) to (h);
        \propag [scalar] (M) to (3);
        \propag [scalar] (4) to (M);
    \end{feynhand}
\end{tikzpicture}
\begin{tikzpicture}[scale=0.95]
    \begin{feynhand}
        \vertex (i) at (-1.5, 0) {$\phi$};
        \vertex (M) at (0, 0);
        \vertex (3) at (1.5, 1) {$\varphi$};
        \vertex (4) at (1.5, -1) {$\varphi$};
        \vertex (h) at (1.5, 0) {$h_{\mu\nu}$};
        \vertex () at (0, -1.5) {};
        \propag [inflaton] (i) to (M);
        \propag [graviton] (M) to (h);
        \propag [scalar] (M) to (3);
        \propag [scalar] (4) to (M);
    \end{feynhand}
\end{tikzpicture}
\caption{Graviton production through the three-body decay $\phi(l)\to\varphi(p)\varphi(q)h(\omega)$. The last diagram contributes nontrivially only in the derivative interaction scenario. \label{fig:diagram}}
\end{figure}

In this section we present the detailed computation of the graviton emission amplitudes, and the phase space as well as the full GW spectrum for inflaton two-body decay, considering both the derivative and nonderivative interactions.

\suppsubsection{Derivative Coupling}\label{sec:App_D}

This subsection presents novel results for decays from a higher-dimension effective operator.  For the derivative operator $\phi\,(\partial_\mu\varphi)(\partial^\mu\varphi)/\Lambda$, graviton emission receives an additional contribution from the four-point contact vertex.
Unlike the nonderivative interaction, this operator sources anisotropic stress directly at the decay vertex and therefore contributes to graviton production.

We label the momenta in the three-body decay as $\phi(l)\to \varphi(p)\,\varphi(q)\,h(\omega)$,
and write the amplitude as the sum of the three external-leg emission diagrams and the contact diagram shown in Fig.~\ref{fig:diagram}.

Using the Feynman rules in Eqs.~\eqref{eq:Feynman_Rules_1}--\eqref{eq:Feynman_Rules_4}, the individual contributions are
\begin{align}
i \mathcal{M}^{\rm (D)}_1 & = \frac{i (p\cdot q)}{\Lambda}\,
\frac{-l_\mu l_\nu}{M_P (l\cdot \omega)}\,\epsilon^{\star\mu\nu}, \\
i \mathcal{M}^{\rm (D)}_2 & = \frac{i\, q\cdot(p+\omega)}{\Lambda}\,
\frac{p_\mu p_\nu}{M_P (p\cdot \omega)}\,\epsilon^{\star\mu\nu}, \\
i \mathcal{M}^{\rm (D)}_3 & = \frac{i\, p\cdot(q+\omega)}{\Lambda}\,
\frac{q_\mu q_\nu}{M_P (q\cdot \omega)}\,\epsilon^{\star\mu\nu}, \\
i \mathcal{M}^{\rm (D)}_4 & = \frac{i}{\Lambda}\,
\frac{(p\cdot q)\eta_{\mu\nu}-2p_\mu q_\nu}{M_P}\,\epsilon^{\star\mu\nu}
= -\frac{i}{\Lambda}\,
\frac{2p_\mu q_\nu}{M_P}\,\epsilon^{\star\mu\nu},
\end{align}
where $\epsilon^{\mu\nu}$ denotes the graviton polarization tensor. $\mathcal{M}_1$ vanishes for a homogeneous inflaton at rest, since the source term $T_{ij}$ is then zero. Moreover, the term proportional to $\eta_{\mu\nu}$ vanishes for an on-shell graviton because $\eta_{\mu\nu}\epsilon^{\mu\nu}=0$. The polarization sum for a massless graviton is~\cite{deAquino:2011ix,Barman:2023ymn}
\begin{equation}
\sum_\text{pol} \epsilon^{\star\mu\nu} \epsilon^{\alpha\beta}
=
\frac12 \left(\hat{\eta}^{\mu\alpha} \hat{\eta}^{\nu\beta} + \hat{\eta}^{\mu\beta} \hat{\eta}^{\nu\alpha} - \hat{\eta}^{\mu\nu} \hat{\eta}^{\alpha \beta}\right),
\end{equation}
with
\begin{equation}
\hat{\eta}_{\mu \nu}
\equiv
\eta_{\mu \nu}
-
\frac{\omega_\mu \bar{\omega}_\nu +\bar{\omega}_\mu \omega_\nu}{\omega\cdot \bar{\omega}},
\end{equation}
where $\omega = (E_\omega, \vec{\omega})$ and $\bar{\omega} = (E_\omega,- \vec{\omega})$. For massless gravitons, $\omega\cdot \bar{\omega} = E_\omega^2+\vec{\omega}^2 = 2E_\omega^2$.

Summing over physical graviton polarizations gives
\begin{widetext}
\begin{align}
|\mathcal{M}^{\rm (D)}|^2 & =
\frac{\left[p^2(\omega \cdot \bar{\omega}) - 2(p\cdot \omega)(p \cdot \bar{\omega})\right]^2
\left[ (p\cdot \omega)\left((p\cdot \omega)+q\cdot \omega\right)- (l\cdot \omega)\left(p\cdot q + 2(p\cdot \omega) + q\cdot \omega\right)
 \right]^2}
{2\Lambda^2 M_P^2 (l \cdot \omega - p\cdot \omega)^2 (p\cdot \omega)^2 (\omega \cdot \bar{\omega})^2}
\nonumber\\
&=
\frac{2(p \cdot \bar{\omega})^2
\left[ (p\cdot \omega)\left((p\cdot \omega)+q\cdot \omega\right)- (l\cdot \omega)\left(p\cdot q + 2(p\cdot \omega) + q\cdot \omega\right)
 \right]^2}
{\Lambda^2 M_P^2 (q\cdot \omega)^2 (\omega \cdot \bar{\omega})^2}
\nonumber\\
&=
\frac{2(p \cdot \bar{\omega})^2
\left[ (p\cdot \omega) (l\cdot \omega)- (l\cdot \omega)\left(p\cdot q + (p\cdot \omega) + l\cdot \omega\right)
 \right]^2}
{\Lambda^2 M_P^2 (q\cdot \omega)^2 (\omega \cdot \bar{\omega})^2}
\nonumber\\
&=
\frac{2(p \cdot \bar{\omega})^2
\left[  (l\cdot \omega)\left(p\cdot q + l\cdot \omega\right)\right]^2}
{\Lambda^2 M_P^2 (q\cdot \omega)^2 (\omega \cdot \bar{\omega})^2}
\nonumber\\
&=
\frac{(l^2)^2}{4\Lambda^2}\times\frac{1}{M_P^2}\times
\left[2\frac{(p \cdot \bar{\omega})^2
(l\cdot \omega)^2}
{(q\cdot \omega)^2 (\omega \cdot \bar{\omega})^2}\right].
\end{align} 
\end{widetext}
In the last step we used momentum conservation, $l-\omega=p+q$, which implies $l\cdot \omega + p \cdot q = (l^2 + q^2 + p^2)/2 = l^2/2$. The factor $(l^2)^2/(4\Lambda^2)$ is equal to $(p\cdot q)^2/\Lambda^2$, namely the squared matrix element for the hard two-body decay through the derivative coupling.

In the inflaton rest frame with $l=(m_\phi,\vec{0})$, the result reduces to
\begin{align}\label{eq:MM_D}
|\mathcal{M}^{\rm (D)}_{\phi\to\varphi\varphi +h}|^2
=
\frac{m_\phi^4}{2\,\Lambda^2 M_P^2}\,
\frac{(1-x)^2}{x^2},
\end{align}
where $x \equiv 2E_\omega/m_\phi$.

\suppsubsection{Nonderivative coupling}\label{sec:App_ND}

For the nonderivative interaction $\mu\,\phi\,\varphi^2/2$, graviton emission arises from attaching the graviton to external legs. This has been studied extensively in the literature, and the results presented in this subsection are included for completeness and for comparison with the derivative case; see e.g.~Ref.~\cite{Bernal:2025lxp}. The amplitudes corresponding to the diagrams in Fig.~\ref{fig:diagram} are
\begin{align}
i \mathcal{M}_1^{\rm (ND)} & = i \mu\,\frac{-l_{\mu} l_{\nu}}{M_P (l \cdot \omega)}\,\epsilon^{\star \mu \nu}, \\
i \mathcal{M}_2^{\rm (ND)} & = i \mu\,\frac{p_{\mu} p_{\nu}}{M_P (p \cdot \omega)}\,\epsilon^{\star \mu \nu}, \\
i \mathcal{M}_3^{\rm (ND)} & = i \mu\,\frac{q_{\mu} q_{\nu}}{M_P (q \cdot \omega)}\,\epsilon^{\star \mu \nu}, \\
i \mathcal{M}_4^{\rm (ND)} & = -i \mu\,\frac{\eta_{\mu \nu}}{M_P}\,\epsilon^{\star \mu \nu}
= -i \mu\,\frac{\epsilon^{\star\mu}{}_{\mu}}{M_P}=0.
\end{align}
The contact contribution $\mathcal{M}_4$, being proportional to $\eta_{\mu\nu}$, vanishes for an on-shell graviton and therefore does not generate graviton emission from the decay vertex.

The total squared matrix element is
\begin{align}
|\mathcal{M}^{\rm (ND)}|^2 & =
\frac{\mu^2 \left[ p^2 (\omega \cdot \bar{\omega})^2 - 2(p\cdot \omega)(p\cdot \bar{\omega})\right]^2 (l \cdot \omega)^2}
{2 M_P^2 \left[(l \cdot \omega) -(p\cdot \omega) \right]^2 (p\cdot \omega)^2 (\omega \cdot \bar{\omega})^2}
\nonumber\\
&=
\frac{2 \mu^2 (p\cdot \bar{\omega})^2 (l \cdot \omega)^2}
{M_P^2 (q \cdot \omega)^2 (\omega \cdot \bar{\omega})^2}
\nonumber \\
&=
\mu^2 \times \frac{1}{M_P^2} \times \left[ 2\frac{(p\cdot \bar{\omega})^2 (l \cdot \omega)^2}
{(q \cdot \omega)^2 (\omega \cdot \bar{\omega})^2}\right],
\end{align}
where the final-state mass has been neglected in the second step. In the inflaton rest frame this simplifies to~\cite{Bernal:2025lxp}
\begin{equation}\label{eq:MM_ND}
|\mathcal{M}_{\phi\to\varphi\varphi +h}^{\rm (ND)}|^2
=
\frac{2 \mu^2}{M_P^2}\,\frac{(1-x)^2}{x^2},
\end{equation}
with $x=2E_\omega/m_\phi$.

It may be surprising that the derivative result Eq.~\eqref{eq:MM_D}  shares
 the same functional form as the nonderivative one Eq.~\eqref{eq:MM_ND} , even though it
contains an additional nonvanishing contact contribution at the decay
vertex.  However, the graviton couples to 
the
energy--momentum flow of the decay.  For a derivative interaction, part of
this flow is localized at the decay vertex itself, and this is why the
contact diagram is present.  After the external-leg diagrams and the
contact diagram are added, the on-shell result is sensitive to the decay
operator only through the hard two-body decay strength.  Therefore the
two expressions coincide after the replacement
$\mu \to m_\phi^2/(2\Lambda)$.

Eqs.~\eqref{eq:MM_D} and \eqref{eq:MM_ND} can be written as
\begin{equation}\label{eq:MM_D_ND}
|\mathcal{M}_{\phi\to\varphi\varphi +h}|^2
=|\mathcal{M}_{\phi\to\varphi\varphi}|^2
\frac{1}{M_P^2} \frac{2}{x^2} \mathcal{F}^{(2)}\,,
\end{equation}
where $ \mathcal{F}^{(2)} = (1-x)^2$. In the soft limit, it becomes
\begin{align}\label{eq:Fsoft_2}
\mathcal{F}^{(2)}_{\text{soft}} =1\,.
\end{align}

\suppsubsection{Phase Space Distribution and Gravitational Wave Spectrum}\label{sec:GW}

With the squared matrix elements in hand, we now compute the resulting gravitational-wave spectrum. To this end, we determine the graviton phase-space distribution $f_h$, which satisfies the Boltzmann equation
\begin{align}
\frac{\partial f_h}{\partial t}
-
H p_h \frac{\partial f_h}{\partial p_h}
=
\mathcal{C},
\end{align}
where $p_h$ denotes the physical graviton momentum and $H$ is the Hubble parameter.

For a generic process
$I_{1}+\cdots+I_{n}\to I_{n+1}+\cdots+I_{n+m}+h$,
the collision term can be written as~\cite{Xu:2025wjq}
\begin{align}
\mathcal{C}
&=\frac{g_{n+m}}{2E_\omega}
\int \left(\prod_{i=1}^{n+m} d\Pi_i\right)
f_{1}\cdots f_{n}
{\cal S}|\mathcal{M}|^{2}\nonumber \\
&
\times (2\pi)^4
\delta^{(4)}
\left(\sum_{i=1}^{n}p_i-\sum_{j=n+1}^{n+m}p_j\right),
\end{align}
where $d\Pi_i=d^3\mathbf{p}_i/[(2\pi)^3 2E_i]$ is the Lorentz-invariant phase-space element, ${\cal S}$ denotes the symmetry factor, and $g_{n+m}=\prod g_i$ accounts for the internal degrees of freedom of the particles involved, excluding the graviton.

Solving Eq.~\eqref{eq:Boltzmann} yields the phase-space distribution $f_h$. Once $f_h$ is known, the GW energy density can be computed as
\begin{align}
\rho_{\text{GW}}
=
g_h
\int \frac{d^3\vec{p}_h}{(2\pi)^3}
p_h f_h
=
g_h
\int \frac{4\pi p_h^3 dp_h}{(2\pi)^3}
f_h,
\end{align}
where $g_h=2$ is the number of graviton polarization states. The present-day GW amplitude is then
\begin{equation}
\Omega_{\text{GW}}(f)
=
\frac{1}{\rho_c}
\frac{d\rho_{\text{GW}}}{d\ln p_h}
=
16\pi^2
\frac{f^4}{\rho_c}
f_h(a_0,2\pi f),
\end{equation}
where $\rho_c\simeq1.05\times10^{-5}h^2\,\text{GeV/cm}^3$~\cite{ParticleDataGroup:2024cfk}
is the critical density, $a_0$ is the scale factor today, and the frequency $f$ is related to the comoving graviton momentum through $p_h(a_0)=2\pi f$.

To illustrate the derivation, we begin with the two-body decay. For the process $\phi(p_1)\to \varphi(p_2)\,\varphi(p_3)\,h(p_4)$, the collision term entering the Boltzmann equation for the graviton distribution is
\begin{align}
\mathcal C(p_4)
&=
\frac{1}{2E_4}
\int d\Pi_1\,d\Pi_2\,d\Pi_3\,
f_1\,
\frac{|\mathcal M_{\phi\to\varphi\varphi h}|^2}{2}\,\nonumber\\
&
\times (2\pi)^4\,
\delta^{(4)}(p_1-p_2-p_3-p_4),
\label{eq:Ch-start}
\end{align}
where $d\Pi_i \equiv d^3\vec p_i/[(2\pi)^3\,2E_i]$, and the factor $1/2$ accounts for the two identical $\varphi$ particles in the final state. Defining the total momentum of the recoiling scalar pair as $Q\equiv p_1-p_4=p_2+p_3$, the phase-space integral over $p_2$ and $p_3$ reduces to the standard two-body phase space,
\begin{align}
\int d\Pi_2\,d\Pi_3\,(2\pi)^4&\delta^{(4)}(Q-p_2-p_3) \nonumber\\
&=
\int d\Phi_2(Q;p_2,p_3)
=
\frac{1}{8\pi},
\end{align}
for massless final-state particles. The inflaton phase-space distribution is
\begin{align}
f_1(\vec p_1)=(2\pi)^3\,n_\phi\,\delta^{(3)}(\vec p_1),
\end{align}
which implies
\begin{align}
\int d\Pi_1\,f_1=\frac{n_\phi}{2m_\phi},
\end{align}
with $n_\phi=\rho_\phi/m_\phi$ the inflaton number density. Substituting these relations into Eq.~\eqref{eq:Ch-start}, and identifying $E_4\equiv E_h$, one finds
\begin{align}
\mathcal C^{(2)}
=
\frac{|\mathcal M_{\phi\to\varphi\varphi h}|^2}{64\pi}\,
\frac{n_\phi}{m_\phi E_h}\,,
\label{eq:Ch}
\end{align}
which reproduces Eq.~(B.4) of Ref.~\cite{Bernal:2025lxp} by a different method. Eq.~\eqref{eq:Ch} can also be written as
\begin{align}
\mathcal C^{(2)} =
\frac{n_\phi \Gamma_{\phi \to  \varphi \varphi}}{M_P^2 m_\phi}\,
\frac{2}{x^3} \mathcal{J}^{(2)}\,, \quad \mathcal{J}^{(2)}=(1-x)^2
\end{align}
 In the soft limit, $\mathcal{J}_{\text{soft}}^{(2)} \to 1$, and the collision term reduces to a form
\begin{align}
\mathcal C^{(2)}_{\rm soft}
\to
\frac{n_\phi \Gamma_{\phi \to \varphi \varphi}}{M_P^2 m_\phi}\,
\frac{2}{x^3},
\label{eq:C-soft-J}
\end{align}
which corresponds to $\mathcal C^{(2)}_{\rm soft}$ in Eq.~\eqref{eq:C_soft_universal}. The decay rates are  $\Gamma_{\phi \to \varphi \varphi} = \frac{m_\phi^3}{128\pi\Lambda^2} $  and $\Gamma_{\phi \to \varphi \varphi} =  \frac{\mu^2}{32\pi m_\phi}$  for the derivative and non-derivative cases, respectively.
To solve Eq.~\eqref{eq:Boltzmann}, it is convenient to introduce the comoving graviton momentum $\tilde p_h \equiv a\,p_h$. In terms of this variable the Boltzmann equation becomes
\begin{align}
\frac{df_h(\tilde p_h)}{dt}
=
\mathcal{C}(\tilde p_h)\,,
\end{align}
where the Hubble dilution term is absorbed. Its formal solution can be written as
\begin{align}\label{eq:fh_int}
f_h(a)
=
\int_{a_I}^{a}
\frac{\mathcal{C}(\tilde p_h)}
{a' H(a')}
\, da',
\end{align}
where $a_I$ denotes the scale factor at the end of inflation, or equivalently the beginning of reheating, and $H_I$ is the Hubble parameter at $a=a_I$. We assume a vanishing initial graviton abundance, $f_h(a_I)=0$.

Using Eq.~\eqref{eq:MM_D} for the derivative case, Eq.~\eqref{eq:Ch}, and the scalings
$H\simeq H_I (a/a_I)^{-3/2}$ and $n_\phi \simeq 3H_I^2 M_P^2/m_\phi\times (a/a_I)^{-3}$ during reheating, Eq.~\eqref{eq:fh_int} yields
\begin{widetext}
\begin{align}\label{eq:fh_arh}
f_h(a_{\rm rh})
&\simeq
\frac{1}{256\pi}
\frac{m_\phi^4\, H_I}{\Lambda^2 p_h^3}
\left(\frac{a_I}{a_{\rm rh}}\right)^{3/2}
\left[1-6\left(\frac{2p_h}{m_\phi}\right) +8\left(\frac{2 p_h}{m_\phi}\right)^{3/2} - 3\left(\frac{2p_h}{m_\phi}\right)^2\right]\nonumber \\
&\simeq
\frac{1}{32\pi}
\frac{m_\phi\, H_I}{\Lambda^2 }
\left(\frac{a_I}{a_{\rm rh}}\right)^{3/2}
\left( \frac{m_\phi}{2 p_h}\right)^3 \simeq \frac{1}{32\pi}
\frac{m_\phi\, H(\arh)}{\Lambda^2 }
\left( \frac{m_\phi}{2 p_h}\right)^3 \simeq \frac{8\,\Gamma_\phi^2}{3\, m_\phi^2} \left( \frac{m_\phi}{2 p_h}\right)^3\,,
\end{align} 
\end{widetext}
evaluated at the end of reheating. After reheating, the collision term becomes negligible as the  inflaton population becomes exponentially small, so that the graviton distribution effectively freezes in and subsequently evolves only through cosmological redshift. The last line of Eq.~\eqref{eq:fh_arh} corresponds to  $f_{h,\text{soft}}$ in Eq.~\eqref{eq:fh_soft_universal}, where we have dropped higher-order terms in $2p_h/m_\phi$, which are small in the soft limit. 

Retaining the higher-order terms in $2p_h/m_\phi$ in Eq.~\eqref{eq:fh_arh}, we  obtain the complete GW spectrum
\begin{widetext}
\begin{align}\label{eq:OGW_full}
\Omega_{\text{GW}} h^2
& \simeq 3.8 \times10^{-18}
\left(\frac{\Trh}{10^{13}\,\text{GeV}}\right)
\left(\frac{m_\phi}{10^{13}\,\text{GeV}}\right)
\left(\frac{f}{10^{10}\,\text{Hz}}\right)  \left[1- 6\left(\frac{f}{f_{\text{th}}}\right)+8 \left(\frac{f}{f_{\text{th}}}\right)^{3/2} - 3\left(\frac{f}{f_{\text{th}}}\right)^2\right],
\end{align}
\end{widetext}
which reproduces the soft spectrum in Eq.~\eqref{eq:Omega_soft_Trh} in the regime $f < f_{\text{th}}$. The expression in brackets reduces to unity as $f\to0$, and vanishes as $f\to f_{\rm th}$. This endpoint corresponds to graviton energies of half of the inflaton mass, $x \to 1$.  From Eq.~\eqref{eq:OGW_full}, we find
\bea
f_{\text{peak}}
&\simeq& 0.21\,f_{\text{th}}\\
&\simeq& 2 \times10^{9}\,\text{Hz}
\left(\frac{m_\phi}{10^{13}\,\text{GeV}}\right)
\left(\frac{10^{13}\,\text{GeV}}{\Trh}\right)\nn
\eea
with peak amplitude
\begin{align}
\Omega^{\text{floor}}_{\text{GW}} h^2
&\simeq 3 \times 10^{-19}
\left(\frac{m_\phi}{10^{13}\,\text{GeV}}\right)^2\,,
\end{align}
which matches the estimate in Eq.~\eqref{eq:floor} in the main text. Eq.~\eqref{eq:OGW_full} is illustrated for the GW spectra from inflaton-two body decays in Fig.~\ref{fig:GW_2body_3_body}.

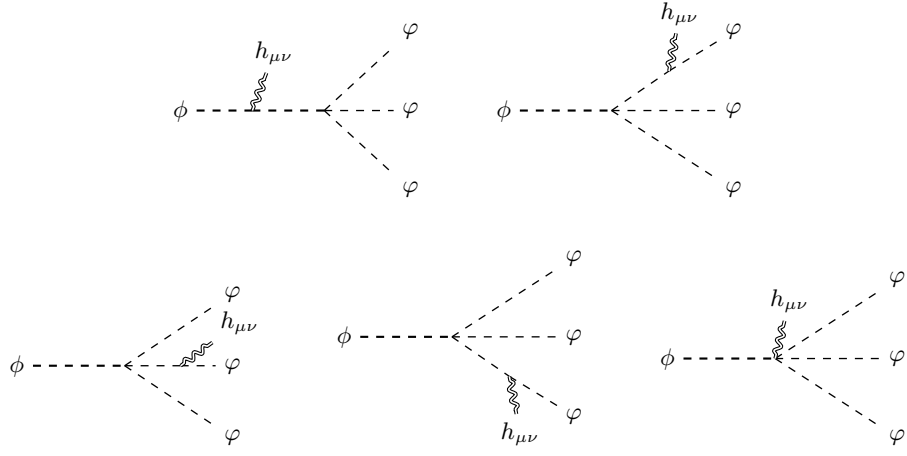
\begin{figure*}[ht]
\centering

\begin{tikzpicture}[scale=0.95]
    \begin{feynhand}
        \vertex (i) at (-1.5,0) {$\phi$};
        \vertex (a) at (-0.5,0);
        \vertex (v) at (0.5,0);
        \vertex (p) at (1.7,1.1) {$\varphi$};
        \vertex (q) at (1.7,0) {$\varphi$};
        \vertex (r) at (1.7,-1.1) {$\varphi$};
        \vertex (h) at (-0.2, 0.8) {$h_{\mu\nu}$};

        \propag [inflaton] (i) to (a);
        \propag [inflaton] (a) to (v);
        \propag [graviton] (a) to (h);
        \propag [scalar] (v) to (p);
        \propag [scalar] (v) to (q);
        \propag [scalar] (v) to (r);
    \end{feynhand}
\end{tikzpicture}
\hspace{0.6cm}
\begin{tikzpicture}[scale=0.95]
    \begin{feynhand}
        \vertex (i) at (-1.5,0) {$\phi$};
        \vertex (v) at (0,0);
        \vertex (a) at (0.8,0.55);
        \vertex (p) at (1.7,1.1) {$\varphi$};
        \vertex (q) at (1.7,0) {$\varphi$};
        \vertex (r) at (1.7,-1.1) {$\varphi$};
        \vertex (h) at (0.95,1.35) {$h_{\mu\nu}$};

        \propag [inflaton] (i) to (v);
        \propag [scalar] (v) to (a);
        \propag [scalar] (a) to (p);
        \propag [scalar] (v) to (q);
        \propag [scalar] (v) to (r);
        \propag [graviton] (a) to (h);
    \end{feynhand}
\end{tikzpicture}

\vspace{0.45cm}

\begin{tikzpicture}[scale=0.95]
    \begin{feynhand}
        \vertex (i) at (-1.5, 0) {$\phi$};
        \vertex (M) at (0, 0);
        \vertex (a) at (0.78, 0);
        \vertex (3) at (1.5, 1) {$\varphi$};
        \vertex (4) at (1.5, 0) {$\varphi$};
        \vertex (5) at (1.5, -1) {$\varphi$};
        \vertex (h) at (1.6, 0.6) {$h_{\mu\nu}$};

        \propag [inflaton] (i) to (M);
        \propag [scalar] (M) to (3);
        \propag [scalar] (M) to (a);
        \propag [scalar] (a) to (4);
        \propag [scalar] (M) to (5);
        \propag [graviton] (a) to (h);
    \end{feynhand}
\end{tikzpicture}
\hspace{0.6cm}
\begin{tikzpicture}[scale=0.95]
    \begin{feynhand}
        \vertex (i) at (-1.5,0) {$\phi$};
        \vertex (v) at (0,0);
        \vertex (a) at (0.8,-0.55);
        \vertex (p) at (1.7,1.1) {$\varphi$};
        \vertex (q) at (1.7,0) {$\varphi$};
        \vertex (r) at (1.7,-1.1) {$\varphi$};
        \vertex (h) at (0.95,-1.35) {$h_{\mu\nu}$};

        \propag [inflaton] (i) to (v);
        \propag [scalar] (v) to (p);
        \propag [scalar] (v) to (q);
        \propag [scalar] (v) to (a);
        \propag [scalar] (a) to (r);
        \propag [graviton] (a) to (h);
    \end{feynhand}
\end{tikzpicture}
\hspace{0.6cm}
\begin{tikzpicture}[scale=0.95]
    \begin{feynhand}
        \vertex (i) at (-1.5,0) {$\phi$};
        \vertex (v) at (0,0);
        \vertex (p) at (1.7,1.1) {$\varphi$};
        \vertex (q) at (1.7,0) {$\varphi$};
        \vertex (r) at (1.7,-1.1) {$\varphi$};
        \vertex (h) at (0.2, 0.8) {$h_{\mu\nu}$};

        \propag [inflaton] (i) to (v);
        \propag [scalar] (v) to (p);
        \propag [scalar] (v) to (q);
        \propag [scalar] (v) to (r);
        \propag [graviton] (v) to (h);
    \end{feynhand}
\end{tikzpicture}

\caption{Graviton production through the four-body decay $\phi(l)\to\varphi(p)\varphi(q)\varphi(r)h(\omega)$. The first four diagrams correspond to graviton emission from the external inflaton and scalar legs, while the last diagram denotes the contact contribution.}
\label{fig:diagram_13h}
\end{figure*}
\suppsection{Graviton Bremsstrahlung from Inflaton $3$-body Decay: Full Spectrum}\label{sec:3body}

In this section, we extend the analysis to inflaton three-body decays and compute the corresponding graviton bremsstrahlung amplitudes. These have not been
studied in previous literature.

\suppsubsection{Emission Amplitude}
We consider the interaction $\mathcal{L}_{\rm int} = \frac{\lambda}{3!}\,\phi\,\varphi^3.$
The process of interest is $\phi(l)\to \varphi(p)\,\varphi(q)\,\varphi(r)\,h(\omega).$
Compared to the two-body decay shown in Fig.~\ref{fig:diagram}, there is now an additional contribution from graviton emission off the external leg $\varphi(r)$, as shown in Fig.~\ref{fig:diagram_13h}. The corresponding matrix elements are
\begin{align}
i \mathcal{M}_1  & = i \lambda \,\frac{-l_{\mu} l_{\nu}}{M_P (l \cdot \omega)}\,\epsilon^{\star \mu \nu}, \\
i \mathcal{M}_2 & = i \lambda \,\frac{p_{\mu} p_{\nu}}{M_P (p \cdot \omega)}\,\epsilon^{\star \mu \nu}, \\
i \mathcal{M}_3 & = i \lambda \,\frac{q_{\mu} q_{\nu}}{M_P (q \cdot \omega)}\,\epsilon^{\star \mu \nu}, \\
i \mathcal{M}_4 & = i \lambda \,\frac{r_{\mu} r_{\nu}}{M_P (r \cdot \omega)}\,\epsilon^{\star \mu \nu}, \\
i \mathcal{M}_5 & = -i \lambda \,\frac{\eta_{\mu \nu}}{M_P}\,\epsilon^{\star \mu \nu}
= -i \lambda \,\frac{\epsilon^{\star\mu}{}_{\mu}}{M_P}=0.
\end{align}
The last contribution vanishes for an on-shell graviton because of the tracelessness condition $\epsilon^{\mu}{}_{\mu}=0$. Summing over the nonvanishing contributions and over graviton polarizations, we obtain
\begin{align}\label{eq:4-body_amplitude}
|\mathcal{M}^2_{\phi \to \varphi \varphi \varphi h}|^2
&=
\lambda^2 \times \frac{1}{M_P^2} \times \frac{2}{x^2} \times \mathcal{F}^{(3)} 
\end{align}
where the kinematic function $\mathcal{F}^{(3)}$ is given by
\begin{widetext}
\begin{align}\label{eq:F}
\mathcal{F}^{(3)} &= \frac{1}{u\,v\,(u+v-x)}
\Big\{
u^{2}\big(v(x+1)^{2}-x(x+z)^{2}\big) 
+u\Big(v^{2}(x+1)^{2}
- vx\big[3x^{2}+2x(y+z+1)-2yz+4y+4z-3\big]\nonumber\\
&\qquad\qquad
+2x^{2}(x+z)(x+y+z-1)\Big)
-x\big[x(-v+y+z-1)-vy+x^{2}\big]^{2}
\Big \}.
\end{align}
\end{widetext}
In the soft limit, the appropriate scaling is not obtained by sending
$x\to 0$ while keeping the remaining variables fixed. Since
$u=2p\cdot\omega/m_\phi^2$ and $v=2q\cdot\omega/m_\phi^2$, one must take the
correlated limit $x\to 0, u=x\,\alpha,v=x\,\beta ,$
with $\alpha$ and $\beta$ fixed. In this limit the function
$\mathcal{F}(x,y,z,u,v)$ approaches a finite angular function,
\begin{widetext}
\begin{align}\label{eq:Fsoft_3}
\mathcal{F}^{(3)}_{\rm soft}(y,z,\alpha,\beta)
=
\frac{1}{\alpha\beta(1-\alpha-\beta)}
\bigg\{
&\left(-1+y-\beta y+z\right)^2
+\alpha^2\left(\beta-z^2\right)
\nonumber\\
&+\alpha\left[
\beta^2+\beta\left(3+2y(-2+z)-4z\right)
+2z(-1+y+z)
\right]
\bigg\}\,,
\end{align}
\end{widetext}
which corresponds to  $\mathcal{F}^{(3)}_{\text{soft}}$  in Eq.~\eqref{eq:universal_soft_decay}.  

\suppsubsection{Collision Term}
We now derive the exact collision term for $\phi(p_1)\to \varphi(p_2)\,\varphi(p_3)\,\varphi(p_4)\,h(p_5),$
which is more involved than Eq.~\eqref{eq:Ch-start} because of the enlarged phase space.

The collision term entering the Boltzmann equation is
\begin{align}
\mathcal C(p_5)
&=
\frac{1}{2E_5}
\int d\Pi_1\,d\Pi_2\,d\Pi_3\,d\Pi_4\,
f_1\,
\frac{|\mathcal M_{\phi\to\varphi\varphi\varphi h}|^2}{3!}\,\nonumber \\
&(2\pi)^4\,
\delta^{(4)}(p_1-p_2-p_3-p_4-p_5),
\label{eq:C-start-13h}
\end{align}
with $d\Pi_i\equiv \frac{d^3\vec p_i}{(2\pi)^3\,2E_i}.$ Using $\int d\Pi_1\,f_1
=
\frac{n_\phi}{2m_\phi},$
Eq.~\eqref{eq:C-start-13h} becomes
\begin{align}
\mathcal C(p_5)
&=
\frac{n_\phi}{4m_\phi E_5}\,
\frac{1}{3!}
\int d\Pi_2\,d\Pi_3\,d\Pi_4\,
|\mathcal M_{\phi\to\varphi\varphi\varphi h}|^2\,
\nonumber \\
&(2\pi)^4\,
\delta^{(4)}(p_1-p_2-p_3-p_4-p_5).
\label{eq:C-afterf-13h}
\end{align}

Define the total momentum of the recoiling three-scalar system,
\begin{align}
Q\equiv p_1-p_5=p_2+p_3+p_4,
\qquad
Q^2=(p_1-p_5)^2.
\end{align}
In the inflaton rest frame,
\begin{align}
x\equiv \frac{2E_5}{m_\phi},
\qquad
Q^2=m_\phi^2-2p_1\cdot p_5=m_\phi^2(1-x).
\label{eq:Q2-def}
\end{align}
Using the recursive factorization
\begin{align}
d\Phi_3(Q;p_2,p_3,p_4)
=
\frac{ds}{2\pi}\,
d\Phi_2(Q;k,p_4)\,
d\Phi_2(k;p_2,p_3)\,
\end{align}
with $k\equiv p_2+p_3\,,
s\equiv k^2$, one can write 
the collision term as
\begin{align}
\mathcal C(x)
=
\frac{n_\phi}{4m_\phi E_5}\,
\frac{1}{3!}
\int d\Phi_3(Q;p_2,p_3,p_4)\,
|\mathcal M_{\phi\to\varphi\varphi\varphi h}|^2.
\label{eq:C-Phi3-13h}
\end{align}

We parameterize the phase space in the rest frame of $Q^\mu$. Let $\theta$ denote the polar angle of $\vec k$ with respect to the graviton direction, and let $(\theta_\ast,\phi_\ast)$ denote the angles of $\vec p_2$ in the rest frame of $k^\mu$. Writing $c\equiv \cos\theta\,,
c_\ast\equiv \cos\theta_\ast,$ the phase-space measure becomes
\begin{align}
d\Phi_3(Q;p_2,p_3,p_4)
=
\frac{Q^2-s}{1024\pi^4 Q^2}\,
ds\,dc\,dc_\ast\,d\phi_\ast,
\label{eq:Phi3-measure-13h}
\end{align}
with integration ranges
\begin{align}
0\le s\le Q^2\,,
-1\le c\le 1\,,
-1\le c_\ast\le 1\,,
0\le \phi_\ast<2\pi.
\end{align}

Substituting Eqs.~\eqref{eq:4-body_amplitude} and \eqref{eq:Phi3-measure-13h} into Eq.~\eqref{eq:C-Phi3-13h}, and using $E_5=xm_\phi/2$, we obtain
\begin{widetext}
\begin{align}
\mathcal C^{(3)}(x)
=
\frac{n_\phi\lambda^2}{6144\pi^4 M_P^2 m_\phi^2}\,
\frac{1}{x^3}
\int_0^{Q^2} ds\,\frac{Q^2-s}{Q^2}
\int_{-1}^{1} dc
\int_{-1}^{1} dc_\ast
\int_0^{2\pi} d\phi_\ast\,
\mathcal F(x,y,z,u,v).
\label{eq:C-exact-s}
\end{align} 
\end{widetext}
Now define the dimensionless invariant
\begin{align}
\hat s\equiv \frac{s}{m_\phi^2},
\qquad
0\le \hat s\le 1-x.
\end{align}
Since $ds=m_\phi^2\,d\hat s$, the factor $m_\phi^2$ from the change of variables cancels the explicit $1/m_\phi^2$ in Eq.~\eqref{eq:C-exact-s}. Using $Q^2=m_\phi^2(1-x)$, the collision term becomes
\begin{align}
\mathcal C^{(3)}(x)
&=
\frac{n_\phi\lambda^2}{1536\pi^3 M_P^2}\,
\frac{(1-x)\mathcal{J}^{(3)}(x)}{x^3}\nonumber\\
& = \frac{n_\phi \Gamma_{\phi \to  \varphi \varphi \varphi}}{M_P^2 m_\phi}\,
\frac{2}{x^3} (1-x)\mathcal{J}^{(3)}(x)\,,
\label{eq:C-exact-J}
\end{align}
where the three-body decay rate is $\Gamma_{\phi \to  \varphi \varphi \varphi} = \frac{\lambda^2\, m_\phi}{3072 \pi^3}$  and we have defined
\begin{widetext}
\begin{align}
\mathcal{J}^{(3)}(x)
\equiv
\frac{1}{4\pi(1-x)^2}
\int_0^{1-x} d\hat s\,(1-x-\hat s)
\int_{-1}^{1} dc
\int_{-1}^{1} dc_\ast
\int_0^{2\pi} d\phi_\ast\,
\mathcal F^{(3)}(x,y,z,u,v).
\label{eq:J-def}
\end{align} 
\end{widetext}
 
\begin{figure}[!ht]
\def\sepf{0.7}
\centering
\includegraphics[scale=\sepf]{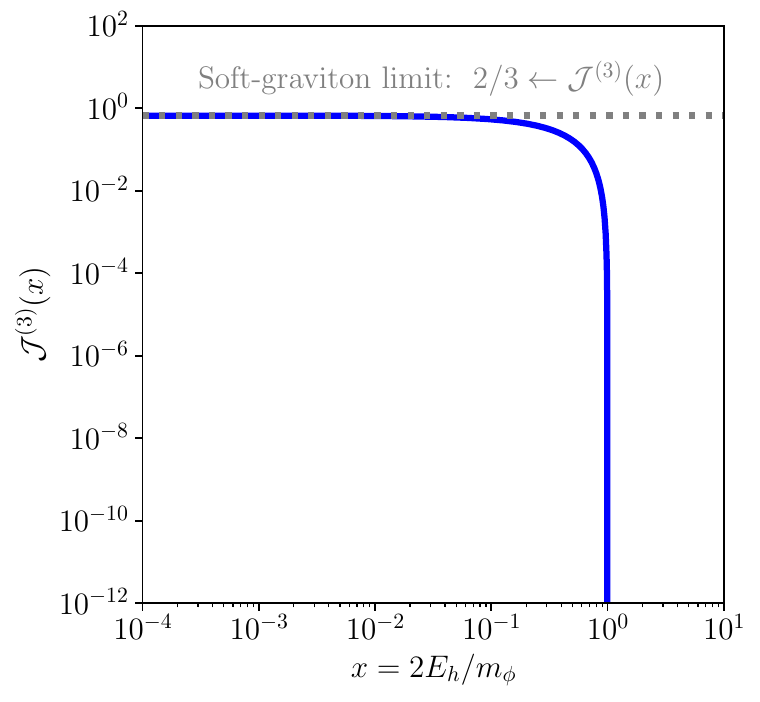}
\caption{Numerical evaluation of $\mathcal{J}^{(3)}(x)$, including the full phase-space integration, as a function of $x$ from $\phi \to \varphi \varphi \varphi +h$.}
\label{fig:Jx}
\end{figure}

It remains to express $y$, $z$, $u$, and $v$ in terms of $(x,\hat s,c,c_\ast,\phi_\ast)$. In particular, one finds
\begin{widetext}
\begin{align}
u
=
\frac{2p_2\cdot p_5}{m_\phi^2}
&=
\frac{x}{4(1-x)}
\Big[
(1-x+\hat s)+(1-x-\hat s)c_\ast
-c\big((1-x-\hat s)+(1-x+\hat s)c_\ast\big)+2\sqrt{\hat s(1-x)(1-c_\ast^2)(1-c^2)}\cos\phi_\ast
\Big],
\label{eq:u-def}
\\[1ex]
v
=
\frac{2p_3\cdot p_5}{m_\phi^2}
&=
\frac{x}{4(1-x)}
\Big[
(1-x+\hat s)-(1-x-\hat s)c_\ast
-c\big((1-x-\hat s)-(1-x+\hat s)c_\ast\big)-2\sqrt{\hat s(1-x)(1-c_\ast^2)(1-c^2)}\cos\phi_\ast
\Big],
\label{eq:v-def}
\end{align}
\end{widetext}
while
\begin{align}
y
&=
\frac12\Big[(1-x+\hat s)+(1-x-\hat s)c_\ast\Big]+u,
\label{eq:y-def}
\\
z
&=
\frac12\Big[(1-x+\hat s)-(1-x-\hat s)c_\ast\Big]+v.
\label{eq:z-def}
\end{align}
Eqs.~\eqref{eq:u-def}--\eqref{eq:z-def}, together with Eq.~\eqref{eq:F}, completely determine the integrand in Eq.~\eqref{eq:C-exact-J}. In Fig.~\ref{fig:Jx}, we show the numerical evaluation of Eq.~\eqref{eq:J-def}, represented by the blue solid line. After integration over the phase space, we find
\begin{align}\label{eq:Jx_soft}
\mathcal{J}^{(3)}_{\text{soft}}(x) \to 2/3  
\end{align}
in the soft limit, as indicated by the gray dotted line.  The numerical computation leading to Eq.~\eqref{eq:Jx_soft}
has been cross-checked using a complementary  method: the eikonal
transverse-traceless phase-space average described in the next section.    

\suppsubsection{Exact Graviton Spectrum beyond Soft Limit}
We now solve the Boltzmann equation for the graviton distribution during reheating using Eq.~\eqref{eq:fh_int}. Our objective is to obtain the  distribution at the end of reheating, $a=a_{\rm rh}$. To this end, we define the dimensionless variable $z\equiv \frac{2p_h(a_{\rm rh})}{m_\phi}
=
\frac{2\tilde p_h}{a_{\rm rh}m_\phi}.$
At an earlier time $a$, the same comoving mode corresponds to $x(a)=\frac{2p_h(a)}{m_\phi}
=
\frac{2\tilde p_h}{a m_\phi}
=
z\,\frac{a_{\rm rh}}{a}.$
Therefore, only times satisfying $x(a)\le 1$ contribute, namely $a\ge z\,a_{\rm rh}$.

Changing variables from $a$ to $x$, one obtains
\begin{align}
f_h(a_{\rm rh},z)
=
\frac{\lambda^2 H_{\rm rh}}{512\pi^3 m_\phi}\,
z^{-3/2}\,
G(z),
\label{eq:fh-final}
\end{align}
where
\begin{align}\label{eq:Gz_full}
G(z)
\equiv
\int_z^1 dx\,
\frac{(1-x)\mathcal{J}(x)}{x^{5/2}}.
\end{align}
Equation~\eqref{eq:fh-final} is the final graviton distribution generated by the decay $\phi\to\varphi\varphi\varphi h$ once perturbative reheating has completed.

In the infrared regime, $J(x)\to 1$ as $x\to 0$, and it then follows that
\begin{align}\label{eq:Gz_soft}
G(z)\xrightarrow[z\ll 1]{}
\int_z^1 dx\,\frac{1-x}{x^{5/2}}
=
\frac{4}{3}+\frac{2}{3}z^{-3/2}-2z^{-1/2}.
\end{align}
so that $f_h(a_{\rm rh},z)\propto z^{-3}.$
This is precisely the same infrared scaling as in the two-body case, as required by soft universality.

Defining the threshold momentum today by $p_{\rm th}
=
\frac{m_\phi}{2}\,\frac{a_{\rm rh}}{a_0}$, it follows that $z
=
\frac{2p_h(a_{\rm rh})}{m_\phi}
=
\frac{p_h(a_0)}{p_{\rm th}}.$
Using Eq.~\eqref{eq:fh-final}, the present-day GW spectrum becomes
\begin{align}
\Omega_{\rm GW}(p_h)
=
\frac{p_h^4}{\pi^2\rho_c}\,
\frac{\lambda^2 H_{\rm rh}}{512\pi^3 m_\phi}\,
\left(\frac{p_h}{p_{\rm th}}\right)^{-3/2}
G\left(\frac{p_h}{p_{\rm th}}\right).
\label{eq:Omega-exact-13h}
\end{align}
Equivalently, in terms of the GW frequency $f=p_h/(2\pi)$,
\begin{align}
\Omega_{\rm GW}(f)
=
\frac{16\pi^2 f^4}{\rho_c}\,
\frac{\lambda^2 H_{\rm rh}}{512\pi^3 m_\phi}\,
\left(\frac{f}{f_{\rm th}}\right)^{-3/2}
G\left(\frac{f}{f_{\rm th}}\right)\,,
\label{eq:Omegaf-exact-13h}
\end{align}
where $f_{\rm th}=\frac{p_{\rm th}}{2\pi}$. This is the  GW spectrum associated with the three-body decay channel shown in Fig.~\ref{fig:GW_2body_3_body}.

The limiting behaviors of Eq.\ (\ref{eq:Omegaf-exact-13h}) are immediately evident. In the soft region, $\Omega_{\rm GW}\propto f$, since $G\propto f^{-3/2}$ according to Eq.~\eqref{eq:Gz_soft}. Near the endpoint $x\to 1$, one has $\Omega_{\rm GW}\to 0$ because the source function $G$ vanishes; cf.~Eq.~\eqref{eq:Gz_full}. Thus the exact spectrum reproduces the universal soft behavior in the infrared, while exhibiting the nonuniversal (but qualitatively similar) hard shapes discussed in the main text.
 
\suppsection{Graviton Bremsstrahlung from Inflaton
$n$-body Decay: Soft Residue and Quadrupole Intuition}\label{sec:TT_ quadrupole}
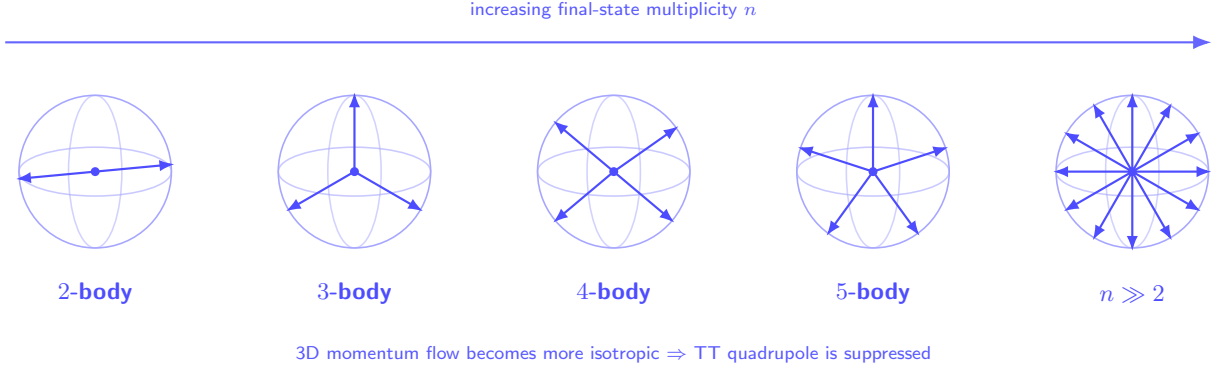
\begin{figure*}[t]
\centering
\makebox[\textwidth][c]{%
\begin{tikzpicture}[
    scale=1.4,
    mom/.style={-{Latex[length=2.0mm]}, line width=0.82pt, blue!70},
    sphere/.style={blue!35, line width=0.60pt},
    lab/.style={font=\sffamily\bfseries\small, blue!70},
    note/.style={font=\sffamily\scriptsize, blue!65}
]

\newcommand{\spheregrid}[1]{
    \draw[sphere] (0,0) circle (#1);
    \draw[sphere, opacity=0.60] (0,0) ellipse (#1 and 0.32*#1);
    \draw[sphere, opacity=0.50] (0,0) ellipse (0.35*#1 and #1);
}

\def\R{0.75}

\begin{scope}[shift={(0,0)}]
    \spheregrid{0.72}
    \fill[blue!70] (0,0) circle (1.2pt);
    \draw[mom] (0,0) -- (\R,0.07);
    \draw[mom] (0,0) -- (-\R,-0.07);
    \node[lab] at (0,-1.15) {$2$-body};
\end{scope}

\begin{scope}[shift={(2.45,0)}]
    \spheregrid{0.72}
    \fill[blue!70] (0,0) circle (1.2pt);
    \draw[mom] (0,0) -- (90:\R);
    \draw[mom] (0,0) -- (210:\R);
    \draw[mom] (0,0) -- (330:\R);
    \node[lab] at (0,-1.15) {$3$-body};
\end{scope}

\begin{scope}[shift={(4.90,0)}]
    \spheregrid{0.72}
    \fill[blue!70] (0,0) circle (1.2pt);
    \draw[mom] (0,0) -- (35:\R);
    \draw[mom] (0,0) -- (140:\R);
    \draw[mom] (0,0) -- (220:\R);
    \draw[mom] (0,0) -- (320:\R);
    \node[lab] at (0,-1.15) {$4$-body};
\end{scope}

\begin{scope}[shift={(7.35,0)}]
    \spheregrid{0.72}
    \fill[blue!70] (0,0) circle (1.2pt);
    \draw[mom] (0,0) -- (90:\R);
    \draw[mom] (0,0) -- (162:\R);
    \draw[mom] (0,0) -- (234:\R);
    \draw[mom] (0,0) -- (306:\R);
    \draw[mom] (0,0) -- (18:\R);
    \node[lab] at (0,-1.15) {$5$-body};
\end{scope}

\begin{scope}[shift={(9.80,0)}]
    \spheregrid{0.72}
    \fill[blue!70] (0,0) circle (1.2pt);

    \foreach \ang in {0,30,...,330}{
        \draw[mom] (0,0) -- (\ang:\R);
    }

    \node[lab] at (0,-1.15) {$n\gg2$};
\end{scope}

\draw[-{Latex[length=2.45mm]}, line width=0.80pt, blue!60]
    (-0.85,1.22) -- (10.55,1.22);

\node[note] at (4.90,1.52)
    {increasing final-state multiplicity $n$};

\node[note] at (4.90,-1.72)
    {3D momentum flow becomes more isotropic $\Rightarrow$ TT quadrupole is suppressed};

\end{tikzpicture}%
}
\caption{
Classical quadrupole intuition for the multiplicity suppression.
For a two-body decay the hard final-state momenta are back-to-back,
selecting a strong axis and giving the largest transverse-traceless
anisotropy.  As the multiplicity increases, the same energy--momentum is
distributed among more directions in three-dimensional phase space,
approaching an isotropic momentum flow whose TT quadrupole is suppressed.
The phase-space average of the soft-graviton source makes this intuition
quantitative, giving $\mathcal J_{\rm soft}^{(n)}=2/n$.
}
\label{fig:quadrupole_intuition}
\end{figure*}

In this section we study soft graviton production from inflaton $n$-body decay with $n\gg 2$. The primary objective is to compute $\mathcal{J}^{(n)}_{\text{soft}}$ shown in
Fig.~\ref{fig:Jn}. Moreover, we present a quadrupole analogy in classical GW emission to  elucidate our numerical results.

Consider the decay of a particle of mass $m_\phi$ at rest into $n$
massless final-state particles: $\phi \to \varphi_1\, \varphi_2 \cdots \varphi_n$. In the soft limit the graviton momentum does
not affect the hard kinematics, so the final momenta satisfy
\begin{equation}\label{eq:phase_space_n}
\sum_{i=1}^n E_i = m_\phi ,
\qquad
\sum_{i=1}^n {\bf p}_i = 0 ,
\qquad
E_i = |{\bf p}_i| \,,
\end{equation}
where $p^{\mu}_i = (E_i, {\bf p}_i)$ denotes the four momentum of the daughter particle.
Equivalently, the total hard four-momentum is
\begin{equation}
\sum_{i=1}^n p_i^\mu = (m_\phi,{\bf 0}) .
\end{equation}

{We generate these hard final-state momenta with the RAMBO algorithm~\cite{Kleiss:1985gy},
which provides an unweighted sampling of the Lorentz-invariant massless
$n$-body phase-space measure.} 
Thus the Monte Carlo average described below is an
 estimate of the Lorentz-invariant phase-space  average of the polarization-summed
soft factor; this is equivalent to doing the phase space integral in Eq.~\eqref{eq:C-start-13h}. {The generated ensemble is rotationally invariant, while each event represents
a definite, generally anisotropic, hard momentum configuration satisfying the
energy--momentum constraints above.}

The relevant soft factor follows from Weinberg's theorem Eq.~\eqref{eq:soft_graviton_theorem}.  For
the emission of a soft graviton with momentum $\omega^\mu$, the amplitude
contains the eikonal tensor
\begin{equation}
S^{\mu\nu}(\omega)
=
\sum_i
\frac{p_i^\mu p_i^\nu}{p_i\cdot \omega} ,
\end{equation}
where the sum runs over the outgoing hard final-state particles, so
$\eta_i=+1$ in Eq.~\eqref{eq:soft_graviton_theorem}.  By rotational
invariance we choose the soft graviton to propagate along the $z$
direction,
\begin{equation}
\omega^\mu = E_\omega(1,0,0,1),
\end{equation}
so that
\begin{equation}
p_i\cdot \omega
=
E_\omega(E_i-p_{z,i}) .
\end{equation}
The common factor $1/E_\omega$ is the universal soft pole.  Since we are
interested here in the finite multiplicity-dependent residue, this common
factor cancels in the normalized ratio below.

For a graviton moving along $z$, the physical transverse-traceless
polarizations lie in the $x$-$y$ plane.  The $+$ polarization projects
onto the $\hat x\hat x-\hat y\hat y$ component, while the $\times$ polarization projects onto
the $\hat x\hat y+\hat y\hat x$ component.  Therefore the finite soft sources for a given
phase-space point are
\begin{align}
T_+
&=
\sum_i
\frac{p_{x,i}^2-p_{y,i}^2}{E_i-p_{z,i}},
\\
T_\times
&=
2\sum_i
\frac{p_{x,i}\,p_{y,i}}{E_i-p_{z,i}} .
\end{align}
We define
\begin{equation}
Q_{\rm soft}^{\rm TT}
=
T_+^2+T_\times^2 .
\end{equation}
Equivalently,
\begin{equation}
Q_{\rm soft}^{\rm TT}
=
\left[
\sum_i
\frac{p_{x,i}^2-p_{y,i}^2}{E_i-p_{z,i}}
\right]^2
+
4
\left[
\sum_i
\frac{p_{x,i}p_{y,i}}{E_i-p_{z,i}}
\right]^2 .
\label{eq:QsoftTT}
\end{equation}
Here ``TT'' denotes the transverse-traceless projection with respect to
the soft graviton direction.  Thus $Q_{\rm soft}^{\rm TT}$ is the
polarization-summed finite soft factor for a given hard final-state
configuration.  The sums over final-state particles are taken before
squaring because the Weinberg soft factor is an amplitude-level source:
the soft graviton couples coherently to the total eikonal stress tensor
of the hard final state.

The form of Eq.~\eqref{eq:QsoftTT} also clarifies the connection with the
usual quadrupole intuition.  A GW propagating along the
$z$ direction is sourced only by the transverse-traceless anisotropy in
the $x$-$y$ plane.  If one ignores the soft propagation factor, the
corresponding quadrupole-like TT strength would be
\begin{equation}
Q_{\rm quad}^{\rm TT}
=
\left[
\sum_i (p_{x,i}^2-p_{y,i}^2)
\right]^2
+
4
\left[
\sum_i p_{x,i}p_{y,i}
\right]^2 .
\end{equation}
This object captures the classical picture that a back-to-back two-body
final state carries a large anisotropic stress, while a many-body final
state is more isotropic on average; this is further illustrated in Fig.~\ref{fig:quadrupole_intuition}.  For soft graviton emission, however,
the radiative source is not the stress $p_i^a p_i^b$, but the
eikonal stress fixed by Weinberg's theorem,
\begin{equation}
p_i^a p_i^b
\quad\longrightarrow\quad
\frac{p_i^a p_i^b}{p_i\cdot\omega}
\propto
\frac{p_i^a p_i^b}{E_i-p_{z,i}},
\qquad a,b=x,y .
\end{equation}
The eikonal denominator changes the energy weighting of each hard particle
and selects the finite soft residue relevant for graviton bremsstrahlung.
Thus Eq.~\eqref{eq:QsoftTT} is the soft-theorem completion of the
quadrupole TT picture.

For each multiplicity $n$, we generate
$N_{\rm MC}=3\times 10^5$
independent massless $n$-body phase-space points satisfying Eq.~\eqref{eq:phase_space_n}.  The phase-space average is defined as
\begin{equation}
\left\langle Q_{\rm soft}^{\rm TT}\right\rangle_n
=
\frac{1}{N_{\rm MC}}
\sum_{a=1}^{N_{\rm MC}}
Q_{\rm soft}^{\rm TT}(a),
\end{equation}
where $a$ labels the Monte Carlo event.  We then normalize to the two-body
result,
\begin{equation}
\mathcal J_{\rm soft}^{(n)}
\equiv
\frac{
\left\langle Q_{\rm soft}^{\rm TT}\right\rangle_n
}{
\left\langle Q_{\rm soft}^{\rm TT}\right\rangle_2
}.
\label{eq:Jsoft_TT_average}
\end{equation}
With this normalization $\mathcal J_{\rm soft}^{(2)}=1$.  This is the
quantity denoted by $J_{\rm soft}^{(n)}$ in the main text.

\begin{figure}[!t]
\def\sepf{0.6}
\centering
\includegraphics[scale=\sepf]{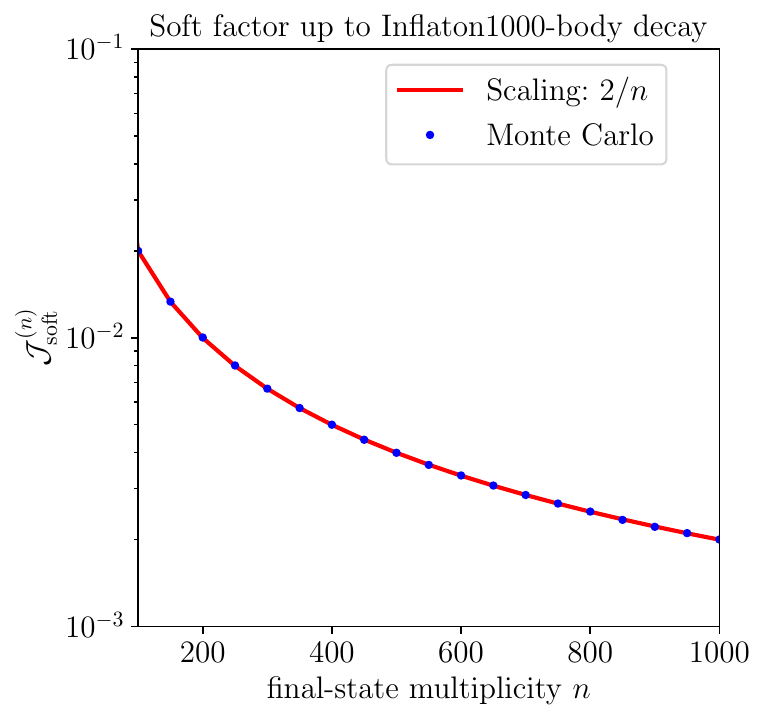}
\caption{Same as Fig.~\ref{fig:Jn} in the main text but for  larger multiplicity  $n \in [10^2, 10^3]$.}
\label{fig:Jn_large_n}
\end{figure}

The numerical results match a scaling 
\begin{equation}
\mathcal J_{\rm soft}^{(n)}
\to 
\frac{2}{n}
\end{equation}
for the multiplicities shown in
Fig.~\ref{fig:Jn} up to $n=100$.  This gives a phase-space interpretation of the
multiplicity suppression, being consistent with the  quadrupole intuition shown in Fig.~\ref{fig:quadrupole_intuition}. We have also checked that our conjecture $\mathcal J_{\rm soft}^{(n)} =\frac{2}{n}$ holds at larger values of $n$, as shown in Fig.~\ref{fig:Jn_large_n}. An analytic proof of this multiplicity law is left to future work.

\bibliography{biblio}
\end{document}